\documentclass[prb,showpacs,twocolumn,aps,superscriptaddress,a4paper]{revtex4}

\usepackage{dcolumn}
\usepackage{amsmath}
\usepackage{graphicx}
\usepackage{latexsym}  
\usepackage{amsfonts}  
\usepackage{amssymb} 
\usepackage{color}
\DeclareGraphicsExtensions{.eps,.pdf,.gif,.jpg,.png}  

\newcommand{\Tr}{ {\rm Tr} \, }
\newcommand{\be}{\begin{equation}}
\newcommand{\ee}{\end{equation}}
\newcommand{\bea}{\begin{eqnarray}} 
\newcommand{\eea}{\end{eqnarray}}

\begin{document}
    
\def\gC{\mbox{\boldmath $C$}}
\def\gZ{\mbox{\boldmath $Z$}}
\def\gR{\mbox{\boldmath $R$}}
\def\gN{\mbox{\boldmath $N$}}
\def\gG{\mbox{\boldmath $G$}}
\def\green{\mbox{\boldmath ${\cal G}$}}
\def\grn{\mbox{${\cal G}$}}
\def\gH{\mbox{\boldmath $H$}}
\def\bA{{\bf A}}
\def\bJ{{\bf J}}
\def\bG{{\bf G}}
\def\bF{{\bf F}}
\def\bH{\mbox{\boldmath $H$}}
\def\bQ{\mbox{\boldmath $Q$}}
\def\bgS{\mbox{\boldmath $\Sigma$}}
\def\bT{\mbox{\boldmath $T$}}
\def\bU{\mbox{\boldmath $U$}}
\def\bV{\mbox{\boldmath $V$}}
\def\bgG{\mbox{\boldmath $\Gamma$}}
\def\bgL{\mbox{\boldmath $\Lambda$}}
\def\ubG{\underline{{\bf G}}}
\def\ubH{\underline{{\bf H}}}
\def\ubQ{\underline{{\bf Q}}}
\def\ubS{\underline{{\bf S}}}
\def\ubg{\underline{{\bf g}}}
\def\ubq{\underline{{\bf q}}}
\def\ubp{\underline{{\bf p}}}
\def\ubgS{\underline{{\bf \Sigma}}}
\def\bge{\mbox{\boldmath $\epsilon$}}
\def\bgD{{\bf \Delta}}

\def\bDelta{\mbox{\boldmath $\Delta$}}
\def\bcalE{\mbox{\boldmath ${\cal E}$}}
\def\bcalF{\mbox{\boldmath ${\cal F}$}}
\def\bcalG{\mbox{\boldmath $G$}}
\def\ubcalG{\mbox{\underline{\boldmath $G$}}}
\def\ubcalA{\mbox{\underline{\boldmath $A$}}}
\def\ubcalB{\mbox{\underline{\boldmath $B$}}}
\def\ubcalC{\mbox{\underline{\boldmath $C$}}}
\def\ubcalg{\mbox{\underline{\boldmath $g$}}}
\def\ubcalH{\mbox{\underline{\boldmath $H$}}}
\def\bcalK{\mbox{\boldmath $K$}}
\def\ubcalK{\mbox{\underline{\boldmath $K$}}}
\def\bcalV{\mbox{\boldmath ${\cal V}$}}
\def\ubcalV{\mbox{\underline{\boldmath $V$}}}
\def\bcalU{\mbox{\boldmath ${\cal U}$}}
\def\ubcalz{\mbox{\underline{\boldmath $z$}}}
\def\bcalQ{\mbox{\boldmath ${\cal Q}$}}
\def\ubcalQ{\mbox{\underline{\boldmath $Q$}}}
\def\ubcalP{\mbox{\underline{\boldmath $P$}}}
\def\bSS{\mbox{\boldmath $S$}}
\def\ubcalS{\mbox{\underline{\boldmath $S$}}}
\def\bff{\mbox{\boldmath $f$}}
\def\bg{\mbox{\boldmath $g$}}
\def\bh{\mbox{\boldmath $h$}}
\def\bk{\mbox{\boldmath $k$}}
\def\bq{\mbox{\boldmath $q$}}
\def\bp{\mbox{\boldmath $p$}}
\def\bt{\mbox{\boldmath $t$}}
\def\ubh{\mbox{\underline{\boldmath $h$}}}
\def\ubt{\mbox{\underline{\boldmath $t$}}}
\def\ubk{\mbox{\underline{\boldmath $k$}}}
\def\ua{\uparrow}
\def\da{\downarrow}
\def\a{\alpha}
\def\b{\beta}
\def\g{\gamma}
\def\G{\Gamma}
\def\d{\delta}
\def\D{\Delta}
\def\e{\epsilon}
\def\ve{\varepsilon}
\def\z{\zeta}
\def\h{\eta}
\def\th{\theta}
\def\vth{\vartheta}
\def\k{\kappa}
\def\l{\lambda}
\def\L{\Lambda}
\def\m{\mu}
\def\n{\nu}
\def\x{\xi}
\def\X{\Xi}
\def\p{\pi}
\def\P{\Pi}
\def\r{\rho}
\def\bgr{\mbox{\boldmath $\rho$}}
\def\s{\sigma}
\def\us{\mbox{\underline{\boldmath $\sigma$}}}
\def\ubgm{\mbox{\underline{\boldmath $\mu$}}}
\def\S{\Sigma}
\def\ubcgS{\mbox{\underline{\boldmath $\Sigma$}}}
\def\t{\tau}
\def\f{\phi}
\def\vf{\varphi}
\def\F{\Phi}
\def\c{\chi}
\def\w{\omega}
\def\W{\Omega}
\def\Q{\Psi}
\def\q{\psi}
\def\de{\partial}
\def\inf{\infty}
\def\ra{\rightarrow}
\def\bra{\langle}
\def\ket{\rangle}
\def\bbra{\langle\langle}
\def\kket{\rangle\rangle}
\def\grad{\mbox{\boldmath $\nabla$}}
\def\no{\bf 1}
\def\ze{\bf 0}
\def\uno{\underline{\bf 1}}
\def\zero{\underline{\bf 0}}

\def\dr{{\rm d}}
\def\bj{{\bf j}}
\def\br{{\bf r}}
\def\bz{\bar{z}}
\def\bart{\bar{t}}

\title{Time-dependent quantum transport with superconducting leads: 
a discrete basis Kohn-Sham formulation and propagation scheme}

\author{Gianluca Stefanucci}
\affiliation{Dipartimento di Fisica, Universit\`a 
di Roma Tor Vergata, Via della Ricerca Scientifica 1, I-00133 Rome, Italy}
\affiliation{European Theoretical Spectroscopy Facility (ETSF)}
\affiliation{Laboratori Nazionali di Frascati, Istituto Nazionale di Fisica
Nucleare, Via E. Fermi 40, 00044 Frascati, Italy}

\author{Enrico Perfetto}
\affiliation{Dipartimento di Fisica, Universit\`a 
di Roma Tor Vergata, Via della Ricerca Scientifica 1, I-00133 Rome, Italy}

\author{Michele Cini}
\affiliation{Dipartimento di Fisica, Universit\`a 
di Roma Tor Vergata, Via della Ricerca Scientifica 1, I-00133 Rome, Italy}
\affiliation{Laboratori Nazionali di Frascati, Istituto Nazionale di Fisica
Nucleare, Via E. Fermi 40, 00044 Frascati, Italy}

\date{\today}

\begin{abstract}
In this work we put forward an exact one-particle framework to study 
nano-scale Josephson junctions out of equilibrium and propose a 
propagation scheme to calculate the time-dependent current in 
response to an external applied bias. 
Using a discrete basis set and Peierls phases for the electromagnetic 
field we prove that the current and pairing densities in a 
superconducting system of interacting electrons can be reproduced in a non-interacting Kohn-Sham 
(KS) system under the influence of different Peierls phases {\em and} 
of a pairing field. 
In the special case of normal systems our 
result provides a formulation of time-dependent current density functional 
theory in tight-binding models.
An extended Keldysh formalism for the non-equilibrium Nambu-Green's 
function (NEGF) is then introduced to calculate the short- and long-time response
of the KS system. The equivalence between the NEGF approach and a 
combination of the static and time-dependent Bogoliubov-deGennes (BdG)
equations is shown. 
For systems consisting of a finite region coupled to ${\cal N}$ 
superconducting semi-infinite leads we numerically solve the static BdG equations 
with a generalized wave-guide approach and their time-dependent 
version with an embedded Crank-Nicholson scheme. To demonstrate the 
feasibility of the propagation scheme we study two paradigmatic 
models, the single-level quantum dot and a tight-binding chain, under 
dc, ac and pulse biases.
We provide a time-dependent picture of single and multiple Andreev 
reflections, show that Andreev bound states can be exploited to 
generate a zero-bias ac current of tunable frequency, and find a 
long-living resonant effect induced by microwave irradiation of 
appropriate frequency.



\end{abstract}

\maketitle

\section{Introduction}

In the last two decades superconducting nanoelectronics has emerged 
as an interdisciplinary field bridging different areas of physics 
like superconductivity, quantum transport and quantum 
computation.\cite{lr.1998,mss.2001,book3} 
For practical applications the reduction of heat losses in 
superconducting circuits constitutes a major advantage over 
semiconductor electronics where a molecular junction is more subject to
 thermal instabilities.\cite{s.1997,ths.2001,dvpl.2002,vsa.2006}

The idea of exploiting atomic-size quantum point contacts or 
quantum dots coupled to superconducting leads 
as quantum bits (QUBIT) has received significant attention both 
theoretically and experimentally.\cite{cw.2008,zsblw.2003,ws.2007,bosht.2007} 
The state of a QUBIT evolves in time according to the Schr\"odinger 
equation for open quantum systems and can be manipulated using 
electromagnetic pulses of the duration of few nano-seconds or even 
faster. Due to the reduced dimensionality and the high speed of the 
pulses these systems can be classified as
ultrafast Josephson nano-junctions (UF-JNJ). 
The microscopic description of the out-of-equilibrium 
properties of an UF-JNJ is not only of importance for their potential 
applications in future electronics but also of 
considerable fundamental interest. 
The quantum nature of the
nanoscale device leads to a sub-harmonic 
gap structure,\cite{rbt.1995,bsw.1995,kdbo.1995,lycldmr.1997,jbsw.1999}  ac 
characteristics,\cite{cmrly.1996,sgw.2007} current-phase 
relation,\cite{b.1991,gsk.1994} etc. that differ substantially from those of 
a macroscopic Josephson junction. Furthermore, there are regimes 
in which the electron-electron scattering inside the device 
plays an important role.\cite{rr.1998,ca.2000,agz.2003,vmrly.2003,gpk.2008}

We here focus on a different relevant aspect of UF-JNJ, namely
the {\em ab initio} description of their short time responses.
Considerable theoretical progresses have been made to construct a 
first-principle scheme of electron transport 
through molecules placed between normal metals. On the contrary, 
despite the recent experimental advances in fabricating 
superconducting quantum point contacts, a 
first-principle approach to superconducting nanoelectronics 
is still missing.
Furthermore, time-dependent (TD) properties 
like the switch on/off time of the current or the response to 
time-dependent ac fields or train pulses has remained largely 
unexplored. There are several 
difficulties related to the construction of a feasible 
time-dependent approach already 
at a mean-field level. The system is open, the electronic 
energy scales are 2-3 orders of magnitude larger than a typical 
superconducting gap, the problem is intrinsically time-dependent (even 
for dc biases), and the possible formation of 
Andreev bound states (ABS) give rise to 
persistent oscillations in the density and current. The time-evolution of localized wave-packets 
scattering across a superconductor-normal interface was explored 
long ago.\cite{k.1969,drmk.1994,jk.2001} More recently the analysis has been 
extended to scattering states  in 
superconductor-device-normal (S-D-N) junctions using the 
wide-band-limit (WBL) approximation\cite{xsw.2007} and in 
superconductor-device-superconductor (S-D-S) junctions by approximating the 
leads with finite size reservoirs.\cite{psc.2009} However, there has been 
no attempt to calculate the response of S-D-S junctions to TD applied 
voltages using truly semi-infinite leads.

In this work we propose a one-particle framework to study 
TD quantum transport in UF-JNJ, construct a suitable 
propagation scheme and apply it to study genuine TD properties 
like the switch on/off of the current, the onset of a Josephson 
regime, ABS oscillations, ac transport and the time-evolution of 
multiple Andreev reflections.

The one-particle framework, described in Section \ref{sysham} 
and \ref{ksscheme}, is an extension of TD 
superconducting density functional theory\cite{wkg.1994} to systems 
with a discrete basis and is built on the mapping from densities to 
potentials proposed by van Leeuwen\cite{vl.1999} and Vignale.\cite{v.2004} 
It is shown that under reasonable assumptions the current density and pairing density of an 
{\em interacting} system perturbed by a TD electromagnetic field can be 
reproduced in a Kohn-Sham system of {\em non-interacting} electrons perturbed by 
a TD electromagnetic {\em and} pairing fields, and that these fields 
are unique. In the special case of normal systems 
such result provides a formulation of TD current density functional 
theory in tight-binding models.

An extended Keldysh formalism for the non-equilibrium Nambu-Green's 
function is introduced in Section \ref{keldsec} and used to calculate the time-dependent 
current, density and pairing density of the Kohn-Sham Hamiltonian.
By adding a vertical imaginary track to the original Keldysh 
contour\cite{d.1984,w.1991,sa.2004} 
we are able to extract the response of the system 
just after the application of the bias (transient regime) and to 
describe the onset of the Josephson regime. 
We also show the equivalence between the equations of motion 
for the Nambu-Green's 
function on the extended contour and 
the combination of the static and TD Bogoliubov-DeGennes equations.

In Section \ref{na} we illustrate a procedure for the calculation of 
the one-particle eigenstates of a system consisting of ${\cal N}$ 
semi-infinite superconducting leads coupled to a finite region $C$.
These states are then propagated in time according to the 
TD Bogoliubov-DeGennes equations using an embedded Crank-Nicholson 
algorithm which reduces to that of Refs. 
\onlinecite{ksarg.2005,spc.2008} in the case of normal leads.
The propagation scheme is unitary (norm conserving) and incorporates 
exactly the transparent boundary conditions.

The feasibility of the method is demonstrated in Section \ref{rts} 
where we calculate the TD current, density and pairing density of S-D-S 
junctions under dc, ac and pulse biases. 
The paradigmatic model with a single atomic level connected to 
a left and right superconducting leads is investigated in detail. We 
provide a time-dependent picture of single and multiple Andreev 
reflections and of the consequent formation of Cooper pairs at 
the interface. We show that the smaller is the bias the longer and 
the more complex is the transient regime. We also study how the 
system relaxes after the bias is switched off. Due to the presence 
of ABS a tiny difference in the switch-off time can cause a large 
difference in the relaxation behavior with {\em persistent oscillations of 
tunable frequency}.
ABS also play a crucial role in microwave ac transport. Tuning the frequency 
of the microwave field according to the ABS energy difference one  
produces a {\em long-living transient resonant effect} in which the amplitude of the ac current 
is about an order of magnitude larger than that of the current out of resonance.
Finally we consider one-dimensional atomic chains coupled 
to superconducting leads. We calculate the TD current density pattern along the 
chain for dc  (ac) biases and 
show  a clear-cut transient scenario 
of the multiple (photon-assisted) Andreev reflections.
A summary of the main findings and an outlook on future 
perspectives are drawn in Section \ref{conc}.

\section{General formulation}

\subsection{Hamiltonian of the system}
\label{sysham}

The Hamiltonian of a system of interacting electrons 
can be written in terms of 
the field operators $\hat{\q}_{\s}(\br)$ ($\hat{\q}^{\dag}_{\s}(\br)$) which destroy 
(create) an electron of spin $\s$ in position $\br$. 
We expand the field 
operators  in some suitable basis of 
localized orbitals $\vf_{m}(\br)$ as $\hat{\q}_{\s}(\br)=\sum_{m}\hat{c}_{m\s}\vf_{m}(\br)$.
Assuming, for simplicity, that the $\vf_{m}$'s are 
orthonormal the $\hat{c}$'s operators obey the anticommutation relations
\be
\{\hat{c}_{m\s},\hat{c}^{\dag}_{n\s'}\}=\d_{\s\s'}\d_{nm}.
\ee
In the presence of an external {\em static}
electromagnetic and pairing field the 
Hamiltonian has the general form
\be
\hat{H}_{0}=\hat{K}_{0}+\hat{\D}_{0}+\hat{\D}^{\dag}_{0}
+\hat{H}_{\rm int}.
\label{ham0}
\ee
The first term is the free-electron part and reads
\be
\hat{K}_{0}=\sum_{\s}\sum_{mn}T_{mn}e^{i\g_{mn}}\hat{c}^{\dag}_{m\s}\hat{c}_{n\s}
\ee
with real symmetric hopping parameters $T_{mn}=T_{nm}$ and real 
antisymmetric phases $\g_{mn}=-\g_{nm}$. The phases 
account for the presence of an external vector potential $\bA(\br)$, 
in accordance with the Peierls prescription. If we use a grid basis 
for the expansion of the field operators with grid points $\br_{m}$ 
then $\g_{mn}=\frac{1}{c}\int_{\br_{n}}^{\br_{m}}d{\bf l}\cdot \bA(\br)$.
The second term in Eq. (\ref{ham0}) represents the pairing field 
operator which couples the pairing density operator to an 
external field and reads
\be
\hat{\D}_{0}=\sum_{m}\D_{m}\hat{c}^{\dag}_{m\ua}\hat{c}^{\dag}_{m\da}.
\ee
We notice that the pairing field $\D_{m}$ is local in the chosen 
basis. This term is usually set to zero since the transition to a 
superconducting state is caused by the interaction part. Our 
motivation to include it at this stage will soon become clear.
The interaction part of the Hamiltonian $\hat{H}_{\rm int}$
contains terms more than quadratic in the $\hat{c}$'s operators. We do not 
specify the form of $\hat{H}_{\rm int}$ which can be any. We, however, require that it 
commutes with the density operator 
$\hat{n}_{m\s}\equiv\hat{c}^{\dag}_{m\s}\hat{c}_{m\s}$
\be
[\hat{H}_{\rm int},\hat{n}_{m\s}]=0,
\quad \forall\, m,\s.
\ee
The above condition is fulfilled on a grid basis as well as 
in tight-binding models with Hubbard-like interactions.

We are interested in the dynamics of the system when an 
extra time-dependent electromagnetic field and pairing potential is switched on at $t=0$.
The pairing potential must here be considered as an independent 
external field.
Since the time-dependent part of the scalar potential can always 
be gauged away we restrict to time-dependent Hamiltonians of the 
form
\be
\hat{H}(t)=\hat{K}(t)+\hat{\D}(t)+\hat{\D}^{\dag}(t)
+\hat{H}_{\rm int},
\label{tdham}
\ee
where 
\be
\hat{K}(t)=\sum_{\s}\sum_{mn}
T_{mn}e^{i\g_{mn}(t)}\hat{c}^{\dag}_{m\s}\hat{c}_{n\s}
\ee
and
\be
\hat{\D}(t)=\sum_{m}
\D_{m}(t)\hat{c}^{\dag}_{m\ua}\hat{c}^{\dag}_{m\da}.
\ee

In 1994 Wacker, K\"ummel and Gross\cite{wkg.1994} put forward a rigorous framework, 
known as TD Density  Functional Theory for Superconductors
(SCDFT), to study the dynamics of a superconducting system in the 
continuum case. 
The continuum Hamiltonian can be obtained from the Hamiltonian  
in Eq. (\ref{tdham}) with the $\vf_{m}$'s  a grid basis in the limit of 
zero spacing.
They proved that given an initial many-body state $|\F_{0}\ket$ 
the current and pairing densities 
evolving under the influence of two different vector potentials 
$\bA$ and $\bA'$ and/or two different pairing 
potentials $\D$ and 
$\D'$ are always 
different. This result renders all observable quantities 
functionals of the current and pairing densities, which can therefore 
be calculated in a one-particle manner.\cite{wkg.1994} 
The original formulation relies
on the assumption that the time-dependent current and pairing 
densities of the interacting Hamiltonian can be reproduced in 
a non-interacting Hamiltonian under the influence of another vector and 
pairing potential, i.e., that the interacting 
$\bA$-$\D$ 
densities are also non-interacting $\bA$-$\D$ representable.
The interacting versus non-interacting representability 
assumption is present also in the original formulation of 
TD Density Functional Theory (DFT) by Runge and Gross\cite{rg.1984} and TD 
Current Density Functional Theory (CDFT) by Ghosh and 
Dhara.\cite{gd.1988}
The representability problem in TDDFT was solved by van 
Leeuwen who 
proved that the TD density of a system with  interaction 
$\hat{H}_{\rm int}$ under the influence of a TD scalar potential $V$ can be reproduced 
in another system with interaction $\hat{H}'_{\rm int}$ under the 
influence of a TD scalar potential $V'$ and that $V'$ is unique.\cite{vl.1999} 
We will refer to such result as the {\em van Leeuwen theorem}.
Taking $\hat{H}'_{\rm int}=0$ the van Leeuwen theorem implies that the TD interacting 
density can be reproduced in a system of non-interacting electrons.
Later Vignale extended the van Leeuwen theorem to solve the 
representability problem in TDCDFT.\cite{v.2004}
In the next section we show that the results by van Leeuwen and 
Vignale can be further extended to solve the representability problem 
in TDSCDFT. The theory is formulated on a discete basis and it
is not limited to pure states, implying that we also have access to 
the finite-temperature domain.

\subsection{The one-particle Kohn-Sham scheme of TDSCDFT}
\label{ksscheme}

Let $\hat{\r}(t)$ be the density matrix at time $t$ of the 
system described by the Hamiltonian in Eq. (\ref{tdham}).
We denote 
by $O(t)\equiv \Tr\{\hat{\r}(t)\hat{O}(t)\}$ the time-dependent 
ensemble average 
of a generic operator $\hat{O}(t)$, where the ``$\Tr$'' symbol 
signifies the trace over a complete set of many-body states.
The average $O(t)$ obeys the equation of motion
\be
\frac{d}{dt}O(t)=\frac{\de}{\de t}O(t)+i\Tr\{\hat{\r}(t)[\hat{H}(t),\hat{O}(t)]\}.
\label{oeqom}
\ee
It is easy to verify that when $\hat{O}(t)$ is the density
operator $\hat{n}_{m}\equiv 
\sum_{\s}\hat{c}^{\dag}_{m\s}\hat{c}_{m\s}$, Eq. (\ref{oeqom}) yields
\be
\frac{d}{dt}n_{m}(t)=\sum_{n}J_{mn}(t)-
4{\rm 
Im}\left[\D_{m}^{\ast}(t)P_{m}(t)e^{-2iT_{mm}t}\right],
\label{conteq}
\ee
where $J_{mn}(t)$ and $P_{m}(t)$ are the expectation value of the 
bond-current operator
\be
\hat{J}_{mn}(t)\equiv
\frac{1}{i}\sum_{\s}\left(
T_{mn}e^{i\g_{mn}(t)}\hat{c}^{\dag}_{m\s}\hat{c}_{n\s}
-{\rm H.c.}
\right)
\label{currdef}
\ee
and pairing density operator 
\be
\hat{P}_{m}(t)\equiv\hat{c}_{m\da}\hat{c}_{m\ua}e^{2i\int_{0}^{t}dt' 
T_{mm}}=
\hat{c}_{m\da}\hat{c}_{m\ua}e^{2i T_{mm}t}.
\ee
Equation (\ref{conteq}) is the proper extension of the continuity equation 
to systems exposed to a pairing field. The term 
$\hat{\D}(t)+\hat{\D}^{\dag}(t)$ acts as if there were TD sources and 
sinks.

Notice that under the gauge transformation 
$\hat{c}_{n\s}\ra e^{i\b_{n}(t)} \hat{c}_{n\s}$ (with $\b_{n}(0)=0$)
the on-site energies change as $T_{mm}\ra T_{mm}-d\b_{m}(t)/dt$ 
while the  phases and the pairing field change according to 
$\g_{mn}(t)\ra\g_{mn}(t)+\b_{m}(t)-\b_{n}(t)$ 
and $\D_{m}(t)\ra \D_{m}(t)\exp[2i\b_{m}(t)]$. 
Therefore the bond-current operator $\hat{J}_{mn}$ and 
pairing density operator $\hat{P}_{m}$ are
gauge invariant. In a grid basis representation with grid points ${\bf r}_{m}$
the phases $\b_{m}(t)$ are the discretized values of 
the scalar function $\L({\bf r}_{m},t)$ which defines the 
gauge-transformed vector potential ${\bf A}$ and scalar potential $V$:
${\bf A}\ra{\bf A}+c\grad \L$ and $V\ra V-\de\L/\de t$.

The equation of motion for the bond-current $J_{mn}(t)$ can 
be cast as follows
\be
\frac{d}{dt}J_{mn}(t)=K_{mn}(t)\frac{d}{dt}\g_{mn}(t)+F_{mn}(t).
\label{eom1}
\ee
The first term in the r.h.s. is exactly $\de J_{mn}(t)/\de t$; the 
operator $\hat{K}_{mn}(t)\equiv\sum_{\s}\left(T_{mn}e^{i\g_{mn}(t)}\hat{c}^{\dag}_{m\s}\hat{c}_{n\s}+{\rm H.c.}
\right)$  is the energy density of the bond $m$-$n$. 
The second term in the r.h.s. is, therefore, the average of 
$\hat{F}_{mn}(t)\equiv i[\hat{H}(t),\hat{J}_{mn}(t)]$, see Eq. (\ref{oeqom}).

The derivation of the equation of motion for the pairing density 
$P_{m}(t)$ is 
also straightforward and leads to 
\bea
\left(\frac{d}{dt}-2iT_{mm}\right)P_{m}(t)&=& i\D_{m}(t)[n_{m}(t)-1]e^{2iT_{mm}t}
\nonumber \\
&+&iG_{m}(t)e^{2iT_{mm}t},
\label{eom2}
\eea
with $\hat{G}_{m}(t)\equiv [\hat{K}(t)+\hat{H}_{\rm int},
\hat{c}_{m\da}\hat{c}_{m\ua}]$.

We now ask the question whether the densities $J_{mn}(t)$ for all 
bonds $m$-$n$ with $T_{mn}\neq 0$ and 
$P_{m}(t)$ can be reproduced in a system with a different interaction 
Hamiltonian $\hat{H}'_{\rm int}$
under the influence of TD phases $\g'(t)$ and pairing 
potential $\D'(t)$ starting from an initial density matrix 
$\hat{\r}'(0)$. 

For the densities to be the same at time $t=0$ we 
have to choose $\hat{\r}'(0)$ and $\g'(0)$ in such a way that 
\be
\Tr\{\hat{\r}'(0)\hat{J}'_{mn}(0)\}=\Tr\{\hat{\r}(0)\hat{J}_{mn}(0)\},
\label{init1}
\ee
\be
\Tr\{\hat{\r}'(0)\hat{P}_{m}(0)\}=\Tr\{\hat{\r}(0)\hat{P}_{m}(0)\}.
\label{init2}
\ee
Notice that in the primed system the bond-current operator 
$\hat{J}'_{mn}$ is different from $\hat{J}_{mn}$ since the phases 
$\g'$ are generally different from $\g$. On the contrary the pairing 
density operator is the same in the two systems.
Equations (\ref{init1},\ref{init2}) define the compatible initial 
configurations of the primed system.  

We answer the above question affirmatively by showing
that given a compatible initial configuration 
$[\hat{\r}'(0),\g'(0)]$
and under reasonable conditions
there exist $\g'(t)$ and $\D'(t)$ for which 
the bond-current and pairing density of the original and
primed system are the same at all times. The formal statement is 
enunciated in the following

{\em Theorem} : Given a compatible initial configuration $[\hat{\r}'(0),\g'(0)]$ 
such that 
\be
K'_{mn}(0)=\Tr\{
\hat{\r}'(0)\sum_{\s}(T_{mn}e^{i\g'_{mn}(0)}\hat{c}^{\dag}_{m\s}\hat{c}_{n\s}+{\rm H.c.}
)\}
\neq 0
\label{init3}
\ee
for all bonds $m$-$n$ with $T_{mn}\neq 0$, and 
\be
n'_{m}(0)=\Tr\{\hat{\r}'(0)\hat{n}_{m}\}\neq 1,
\label{init4}
\ee
which  implies that  at time $t=0$ none of the orbitals $\vf_{m}$ are half 
filled in the primed system,
there exist a unique set of continuous phases $\g'(t)$ and pairing potential 
$\D'(t)$ that reproduce in the primed system the densities 
$J_{mn}(t)$ and $P_{m}(t)$ of the original system.

{\em Remarks} : Before presenting the proof of the Theorem we discuss 
few relevant implications. 
(1) If the original system is a superconducting system with an 
attractive interaction $\hat{H}_{\rm int}$ and a vanishing pairing 
field, i.e., $\hat{\D}=0$, the theorem 
implies that the bond-currents and pairing densities 
can be reproduced in a system of non-interacting 
electrons, i.e., $\hat{H}'_{\rm int}=0$ perturbed by  
TD phases $\g'$ {\em and} pairing field $\D'$. In the following we will refer to such non-interacting 
system as the Kohn-Sham (KS) system and to the TD perturbation as the 
KS phases and KS pairing potential.
In Section \ref{na} we describe how to perform the 
time-evolution of such KS systems for geometries relevant 
to quantum transport.
(2) For interacting systems with $\D=0$ and initially 
in equilibrium in the absence of electromagnetic fields 
the phases $\g(0)=0$ and hence $J_{mn}(0)=0$ for all bonds. 
In the KS system a possible compatible initial configuration is 
therefore $\g'(0)=0$ and $\hat{\r}'(0)$ such that 
the expectation value of the one-particle density matrix 
$n'_{mn}(0)=\sum_{\s}\Tr\{\hat{\r}'(0)\hat{c}^{\dag}_{m\s}\hat{c}_{n\s}\}$ is real.
For such initial configurations the condition (\ref{init3}) becomes 
$n'_{mn}(0)\neq 0$ for all bonds $m$-$n$ with $T_{mn}\neq 0$.
(3) If we ask the question whether only the bond-currents $J_{mn}(t)$ 
of a system with Hamiltonian (\ref{tdham}) and {\em zero pairing 
field}, i.e., $\D=0$, 
can be reproduced in a system with {\em zero pairing 
field}, i.e., $\D'=0$, and different interactions 
$\hat{H}'_{\rm int}$ under 
the influence of different phases $\g'$ starting from 
some initial density matrix $\hat{\r}'(0)$, the answer is affirmative 
provided that $\hat{\r}'(0)$ and $\g'(0)$ fulfill Eqs. (\ref{init1},\ref{init3}).
This corollary extends TDCDFT to tight-binding models using the 
Peierls phases as the basic KS fields and lays down the basis for a  
density functional TD theory in discrete systems.\cite{notetddft}

We conclude this Section with the proof of the Theorem.

{\em Proof} :
The current and pairing densities of the primed system obey the  
equations of motion (\ref{eom1},\ref{eom2}) 
with $K_{mn}(t)\ra K'_{mn}(t)$, $F_{mn}(t)\ra F'_{mn}(t)$ and 
$n_{m}(t)\ra n'_{m}(t)$, $G_{m}(t)\ra G'_{m}(t)$. Therefore, for a generic time $t$ the 
densities  of the two  systems are the same provided that 
\bea
K'_{mn}(t)\frac{d }{dt}\g'_{mn}(t)&=&
K_{mn}(t)\frac{d }{dt}\g_{mn}(t)
\nonumber \\ &+&F_{mn}(t)-F'_{mn}(t),
\label{eqap}
\eea
\bea
[n'_{m}(t)-1]\D'_{m}(t)&=&[n_{m}(t)-1]\D_{m}(t)
\nonumber \\ 
&+&G_{m}(t)-G'_{m}(t).
\label{eqdp}
\eea
A discussion on the existence and uniqueness 
of the solution for the coupled Eqs. (\ref{eqap}-\ref{eqdp}) is 
rather complicated since the dependence on the phases $\g'$
and potentials $\D'$ in $F'$ and $G'$ enters implicitly via
the TD density matrix $\hat{\r}'(t)$. To proceed further we then follow 
the approach of Vignale and assume that the time-dependent phases and 
pairing potentials 
and hence all expectation values are analytic functions of 
time around $t=0$.\cite{v.2004} Expanding all quantities in Eqs. 
(\ref{eqap}-\ref{eqdp}) in their Taylor series and equating the 
coefficients with the same power of $t$ we obtain 
\bea
(l+1)K'^{(0)}_{mn}\g'^{(l+1)}_{mn}
&=&-\sum_{k=0}^{l-1}(k+1)K'^{(l-k)}_{mn}
\g'^{(k+1)}_{mn}
\nonumber \\
&+&
\sum_{k=0}^{l}(k+1)K_{mn}^{(l-k)}
\g_{mn}^{(k+1)}
\nonumber \\
&+&F'^{(l)}_{mn}-F_{mn}^{(l)},
\label{recura}
\eea
\bea
[n'^{(0)}_{m}-1]\D'^{(l)}_{m}
&=&-\sum_{k=0}^{l-1}n'^{(l-k)}_{m}\D'^{(k)}_{m}
\nonumber \\
&+&\sum_{k=0}^{l}n^{(l-k)}_{m}\D^{(k)}_{m}-\D_{m}^{(l)}
+G'^{(l)}_{m}-G^{(l)}_{m},
\nonumber \\
\label{recurd}
\eea
where for a generic analytic function $f(t)$ 
we defined $f^{(l)}$ as the $l$-th coefficient of the Taylor expansion. 
We now show that Eqs. (\ref{recura}-\ref{recurd}) constitute a 
set of recursive relations to calculate all $\g'^{(l)}$ and $\D'^{(l)}$ 
once all $\g'^{(k)}$ and $\D'^{(k)}$ are known for $k<l$.
We first observe that the $l$-th derivative of the density matrix
$\hat{\r}'(t)$ in $t=0$ depends at most on the $(l-1)$ derivative 
of $\g'$ and $\D'$ since 
$i\frac{d}{dt}\hat{\r}'(t)=[\hat{H}'(t),\hat{\r}'(t)]$. 
The quantity $F'_{mn}$ 
depends on $(\g',\D')$ implicitly through $\hat{\r}'(t)$ and 
explicitly through the commutator $[\hat{H}'(t),\hat{J}'_{mn}(t)]$.
Since the $l$-th derivative of the commutator 
depends on all $(\g'^{(k)},\D'^{(k)})$
with $k\leq l$ the quantity $F'^{(l)}_{mn}$ 
is a function of  $(\g'^{(k)},\D'^{(k)})$ with $k\leq l$. On the 
contrary, the quantities $K'$,
$G'$ depend implicitly on $(\g',\D')$  through $\hat{\r}'(t)$ but 
they explicitly depend only on $\g'$, i.e., there is no 
explicit dependence on the pairing potential $\D'$. We therefore 
conclude that $K'^{(l)}$ and $G'^{(l)}$ depend
on the $\g'^{(k)}$ with 
$k\leq l$ and on $\D'^{(k)}$ with $k<l$. 
Finally, from Eq. (\ref{conteq}) we see that the $l$-th derivative of 
the density $n'_{m}(t)$ depends at most on the $l-1$ derivative of 
$\g'$ and $\D'$. The table below summarizes the dependency of the 
various quantities on the order of the derivatives of $\g'$ and $\D'$
\be
\begin{array}{c|c|c|c|c}
  &  F'^{(l)} & K'^{(l)} & G'^{(l)} & n'^{(l)} \\
  \hline
  & & & & \\
  \{\g'^{(k)}\} & k\leq l & k\leq l & k\leq l & k< l \\
   & & & & \\
  \hline
  & & & & \\
  \{\D'^{(k)}\} & k\leq l & k<l & k<l & k<l 
\end{array}
\ee
    
From the above considerations it follows that Eq. (\ref{recurd}) 
with $l=0$ can be used to determine $\D'^{(0)}$ since the r.h.s.  
depends only on $\g'^{(0)}=\g'(0)$ and from Eq. (\ref{init4}) the 
prefactor $[n'^{(0)}_{m}-1]\neq 0$.
Having $\D'^{(0)}$ we can easily calculate $\g'^{(1)}$ from Eq. (\ref{recura}) 
with $l=0$ since the r.h.s. depends only on $\g'^{(0)}$ and 
$\D'^{(0)}$ and from Eq. (\ref{init3})
$K'^{(0)}_{mn}\neq 0$.
With $\g'^{(1)}$, $\g'^{(0)}$ and $\D'^{(0)}$ we can use Eq. 
(\ref{recurd}) with $l=1$ to extract $\D'^{(1)}$, then Eq. 
(\ref{recura}) with $l=1$ to extract $\g'^{(2)}$ and so on and so 
forth.

\subsection{Keldysh-Green's function in the Nambu space}
\label{keldsec}

\subsubsection{Keldysh contour}

We now specialize to interacting systems which are initially in equilibrium at 
temperature $T=1/\b$ and chemical potential $\m$; such initial 
configurations are the relevant ones in quantum transport 
experiments, see Section \ref{qtappl}.\cite{geninit}
From static SCDFT\cite{ogk.1988} we can choose the initial density 
matrix of the KS system as the thermal density matrix of a system described 
by the equilibrium Hamiltonian (\ref{ham0}) with 
$\hat{H}_{\rm int}=0$ and KS phases $\g$ and pairing potentials 
$\D$, and from the results of the previous section we know that 
such KS system can reproduce the TD bond-currents and pairing 
densities of the interacting system if perturbed by TD KS phases 
$\g(t)$ and pairing potentials $\D(t)$.
Denoting by $\hat{H}_{s}(t)=\hat{K}(t)+\hat{\D}(t)+\hat{\D}^{\dag}(t)$ 
the TD Hamiltonian and by 
$\hat{\r}_{s}(t)$ the TD density matrix of the KS system we then have
\be
\hat{\r}_{s}(t)=\frac{1}{\cal Z}\hat{S}_{s}(t)e^{-\b(\hat{H}_{s}-\mu \hat{N})}
\hat{S}_{s}^{\dag}(t)
\ee
where ${\cal Z}=\Tr\{e^{-\b(\hat{H}_{s}-\mu \hat{N})}\}$ is the 
partition function and $\hat{S}_{s}(t)$ is the KS evolution operator
to be determined from 
$i\frac{d}{dt}\hat{S}_{s}(t)=\hat{H}_{s}(t)\hat{S}_{s}(t)$ with boundary 
condition $\hat{S}_{s}(0)=1$. The Hamiltonian 
$\hat{H}_{s}=\hat{H}_{s}(0)$ is the equilibrium KS Hamiltonian while 
$\hat{N}$ is the total number of particles operator. 
It is worth to notice that in general $[\hat{H}_{s},\hat{N}]\neq 0$ 
due to the presence of the pairing field.
The TD expectation value $O_{s}(t)$ of a 
generic operator $\hat{O}(t)$ is in the KS system given 
by\cite{d.1984,w.1991,sa.2004,vldsavb.2006}
\be
O_{s}(t)=\Tr\{\hat{\r}_{s}(t)\hat{O}(t)\}
\equiv\bra T_{\rm K}\left\{
\hat{O}(z=t_{\pm}) \right\}\ket
\label{genev}
\ee
where we have introduced the short hand notation
\begin{equation}
\bra T_{\rm K}\{\ldots\}\ket=\frac{\Tr\left[  
T_{\rm K}\left\{e^{-i\int_{\g_{\rm K}}\dr 
\bz\,\hat{H}_{\m,s}(\bz)}
\ldots\right\}\right]}
{\Tr\left[ T_{\rm K}\left\{e^{-i\int_{\g_{\rm K}}\dr 
\bz\,\hat{H}_{\m,s}(\bz)}\right\}\right]}.
\label{vrgsc}
\end{equation}
\begin{figure}[t]
\centering
\includegraphics[width=0.48\textwidth]{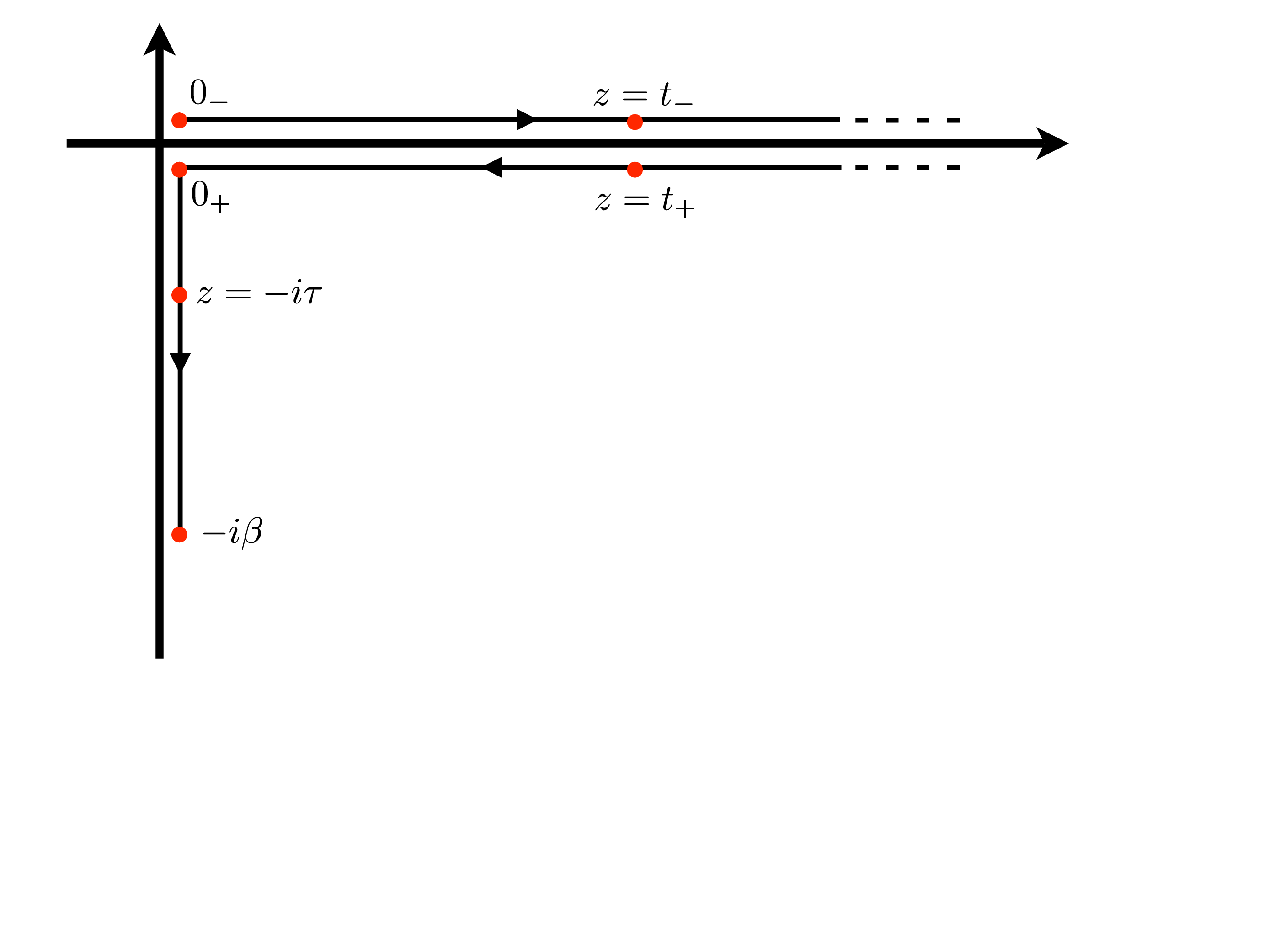}
\caption{The Keldysh contour $\g_{\rm K}$ described in the main text. 
The contour variable $z=t_{-}/t_{+}$ denotes a point on the 
upper/lower branch at a distance $t$ from the origin while 
$z=-i\t$ denotes a point on the imaginary track at a distance $\t$ 
from the origin. In the figure we also illustrate the points 
$0_{-}$ (earliest point on $\g_{\rm K}$), $0_{+}$ and $-i\b$ (latest 
point on $\g_{\rm K}$).}
\label{keldyshc}
\end{figure}
In the above equation $\g_{\rm K}$ is the Keldysh contour\cite{k.1965} illustrated in Fig. 
\ref{keldyshc} which is an oriented contour composed by an 
upper branch going from $0$ to $\inf$, a lower branch going from 
$\inf$ to $0$ and a purely imaginary (thermal) segment going from $0$ to 
$-i\b$. The operator $T_{\rm K}$  is the contour 
ordering operator and move operators with later contour variable to 
the left (an extra minus sign has to be included for odd permutations 
of fermion fields). 
Finally $\hat{H}_{\m,s}(\bz=\bar{t}_{\pm})=\hat{H}_{s}(\bar{t})$ 
where the contour points $\bar{t}_{-}/\bar{t}_{+}$ lie on the 
upper/lower branch at a distance $\bar{t}$ from the origin
while for $\bz$ on the thermal segment 
$\hat{H}_{\m,s}(\bz=-i\t)=\hat{H}_{s}-\mu\hat{N}$. Thus, the 
denominator in Eq. (\ref{vrgsc}) is simply the partition function 
${\cal Z}$.
In Eq. (\ref{genev}) the variable $z$ on the 
contour can be taken either on the upper ($t_{-}$) or lower ($t_{+}$) 
branch at a distance $t$ from the origin. 

\subsubsection{Keldysh-Nambu-Green's function}

The KS expectation value $O_{s}(t)$ of an 
operator $\hat{O}(t)$ is in general different from the 
expectation value $O(t)$ produced by the original system. However if 
$\hat{O}(t)$ is the KS bond-current operator or the pairing density 
operator the average over the KS system yields exactly the  
bond-current and pairing density of the original system.
It is therefore convenient to introduce 
the non-equilibrium Nambu-Green's functions (NEGF) from which the
expectation value of any one-particle operator can be extracted. 
A further reason for us to introduce the NEGF
is that the equilibrium and time-dependent 
Bogoliubov-deGennes equations can be elegantly derived from them,
thus illustrating the equivalence between the NEGF 
and the Bogoliubov-deGennes formalisms.
The normal and anomalous components of the NEGF 
are defined according to\cite{n.1960}
\begin{eqnarray}
\bG_{\s,mn}(z;z')&=&
\frac{1}{i}\bra T_{\rm K}\left\{\hat{c}_{m\s}(z)\hat{c}^{\dag}_{n\s}(z') \right\}\ket,
\label{gn}
\\ \nonumber \\
\bF_{mn}(z;z')&=&
\frac{1}{i}\bra T_{\rm K}\left\{\hat{c}_{m\da}(z)\hat{c}_{n\ua}(z') \right\}\ket,
\label{gan}
\\ \nonumber \\
\overline{\bF}_{mn}(z;z')&=&
-\frac{1}{i}\bra T_{\rm K}\left\{\hat{c}^{\dag}_{n\ua}(z')\hat{c}^{\dag}_{m\da}(z) \right\}\ket,
\label{ganc}
\end{eqnarray}
where $z,z'$ run on the Keldysh contour 
$\g_{\rm K}$.\cite{d.1984,w.1991,rs.1986,vldsavb.2006} 
The $\hat{c}$ operators carry a dependence on the $z$ variable;
such dependence simply specifies their position along the contour 
so to have a well defined action of $T_{\rm K}$.\cite{vldsavb.2006}
The TD bond-current and pairing density can be expressed in 
terms of $\bG_{\s}(z;z')$ and 
$\bF(z;z')$ as 
\be
J_{mn}(t)=-\sum_{\s}\left(T_{mn}e^{i\g_{mn}(t)}
\bG_{\s,nm}(t_{-};t_{+})+{\rm H.c.}
\right),
\label{avecurr}
\ee
\be
P_{m}(t)=i\bF_{mm}(t_{+};t_{-})e^{2iT_{mm}t}.
\label{avepden}
\ee

\subsubsection{Equations of motion}

The NEGF of the KS system obey the following  
equations of motion
\begin{eqnarray}
\left\{i\frac{\overrightarrow{d}}{d z}\,\uno
-\ubH_{\m}(z)\right\}\ubG(z;z')=\uno\d(z-z'),
\label{lem}
\\ \nonumber \\ 
\ubG(z;z')\left\{-i\frac{\overleftarrow{d}}{d 
z'}\,\uno-\ubH_{\m}(z')\right\}=\uno\d(z-z'),
\label{rem}
\end{eqnarray}
where all underlined quantities are $2\times 2$ matrices in the  Nambu 
space with matrix elements $\uno_{mn}=\left[\begin{array}{cc} \d_{mn} & 0 \\ 0 & 
\d_{mn} \end{array}\right]$ and
\be
\ubG_{mn}(z;z')=
\left[
\begin{array}{cc}
    \bG_{\ua,mn}(z;z') & -\bF_{nm}(z';z) \\ \\
    \overline{\bF}_{mn}(z;z') & -\bG_{\da,nm}(z';z)
\end{array}
\right],
\label{kgfcomp}
\ee
\begin{equation}
\ubH_{\m,mn}(z)=\left[
\begin{array}{cc}
    K_{\m,mn}(z) & \d_{mn}\D_{m}(z) \\ \\
    \d_{mn}\D^{\ast}_{m}(z) & -K_{\m,nm}(z)
\end{array}
\right].
\end{equation}
The matrix elements of $\ubH_{\m}(z)$ are 
\be
\left\{
\begin{array}{l}
K_{\m,mn}(t_{\pm})=T_{mn}e^{i\g_{mn}(t)}
\\
\D_{m}(t_{\pm})=\D_{m}(t)
\end{array}
\right.
\label{hmateler}
\ee
for $z=t_{\pm}$ on the horizontal 
branches and 
\be
\left\{
\begin{array}{l}
K_{\m,mn}(-i\t)=T_{mn}e^{i\g_{mn}}-\m\d_{mn}
\\
\D_{m}(-i\t)=\D_{m}
\end{array}
\right.
\label{hmatelei}
\ee
for $z=-i\t$ on the imaginary track. 
Since $\ubH_{\m}(-i\t)$ is independent of $\t$
we write $\ubH_{\m}(-i\t)=\ubH_{0}-\mu\us$ with 
$\us_{mn}=\s_{z}\uno_{mn}$ and $\s_{z}$ the third Pauli matrix.

In the next Section we show that the solution of the equations of 
motion is equivalent to first solve the static Bogoliubov-deGennes 
(BdG) equations and then their TD version.

\subsubsection{Keldysh components and Bogoliubov-deGennes equations}

We introduce the left and right contour evolution matrices 
$\ubS^{R/L}(z)$ which satisfy 
\bea
i\frac{d}{d 
z}\ubS^{R}(z)&=&\ubH_{\m}(z)\ubS^{R}(z),
 \\
-i\frac{d}{d z'}\ubS^{L}(z')&=&\ubS^{L}(z')\ubH_{\m}(z'),
\eea
with boundary conditions $\ubS^{R/L}(0_{-})=\uno$. 
The most general solution of the equations of 
motion (\ref{lem},\ref{rem}) can then be written as 
\begin{equation}
\ubG(z;z')=\ubS^{R}(z)
\left[
\theta(z;z')\ubG^{>}+\theta(z';z)\ubG^{<}\right]
\ubS^{L}(z'),
\label{scsol}
\end{equation}
with $\ubG^{>}-\ubG^{<}=-i\uno$ and
the contour Heaviside function $\theta(z;z')=1$ if $z$ is later than $z'$ and 
zero otherwise. Equation (\ref{scsol}) is a solution 
for all matrices
$\ubG^{>}=-i\uno+\ubG^{<}$. In order to determine $\ubG^{>}$ or 
$\ubG^{<}$ we use the boundary conditions
\bea
\ubG(0_{-};z')&=&-\ubG(-i\b;z'),
\label{bcsc1} \\
\ubG(z;0_{-})&=&-\ubG(z;-i\b),
\label{bcsc2}
\eea
which follow directly from the definitions 
(\ref{gn}-\ref{ganc}) of the NEGF.
Using Eq. (\ref{scsol}) one finds
$\ubG(0_{-};z')=\ubG^{<}\ubS^{L}(z')$ and 
$\ubG(-i\b;z')=\ubS^{R}(-i\b)\ubG^{>}\ubS^{L}(z')$
from which we conclude that 
\be
\ubG^{<}=-\ubS^{R}(-i\b)\ubG^{>}.
\ee
Similarly, from Eq. (\ref{bcsc2}) one finds
\be
\ubG^{>}=-\ubG^{<}\ubS^{L}(-i\b).
\ee
Exploiting the fact that $\ubH_{\m}(-i\t)=\ubH_{0}-\m\us$ 
is constant along the imaginary track one readily realizes that
$\ubS^{R/L}(-i\b)=\exp[\pm\b(\ubH_{0}-\m\us)]$ and hence
\begin{equation}
\ubG^{<}=\frac{i}{\uno+\exp[\b(\ubH_{0}-\m\us)]}.
\label{g><scf}
\end{equation}
From the exact solution (\ref{scsol}) we can extract any observable 
quantity at times $t\geq 0$ and not only its limiting behavior at 
$t\ra\inf$. Below we calculate the different components of 
the NEGF.

We introduce the eigenstates $\Q_{q}$, with eigenenergies 
$E_{q}$, of the matrix $\ubH_{0}-\m\us$. 
The vector 
$\Q_{q}=[u_{q},v_{q}]$ is a two-dimensional vector in the Nambu space
and, by definition, satisfies the eigenvalue problem
\be
\sum_{n}T_{mn}e^{i\g_{mn}}u_{q}(n)+\D_{m}v_{q}(m)=
(E_{q}+\m)u_{q}(m),
\label{es1}
\ee
\be
-\sum_{n}T_{nm}e^{i\g_{nm}}v_{q}(n)+\D^{\ast}_{m}u_{q}(m)=
(E_{q}-\mu)v_{q}(m).
\label{es2}
\ee
Due to the presence of the pairing field the components $u_{q}$ and 
$v_{q}$ are coupled and the eigenstates $\Q_{q}$ are a mixture of 
one-particle spin-up electron states and spin-down hole states. We 
will refer to the eigenstates $\Q_{q}$ as {\em bogolons}.
The above equations have the structure of the static BdG equations 
which follow from the BCS approximation.\cite{a.1964,dg.1966} 
In our case Eqs.  (\ref{es1},\ref{es2}) follow from 
SCDFT\cite{ogk.1988} and  
therefore yield the exact equilibrium bond-current and pairing 
density provided that the exact KS phases and pairing fields are 
used.  

Inserting the complete set of eigenstates in Eq. (\ref{scsol}) and taking into 
account Eq. (\ref{g><scf}) we find the following expansion for the 
NEGF
\bea
\ubG(z;z')=i\sum_{q}\,&&\!\!\!\!\!\!\ubS^{R}(z)\Q_{q}
\left[
\theta(z;z')f^{>}(E_{q})
\right.
\nonumber \\
&&\!\!\!\!\!\!\left.+\theta(z';z)f^{<}(E_{q})\right]\Q_{q}^{\dag}
\,\ubS^{L}(z'),
\label{scsolex}
\eea
where $f^{<}(\w)=1/[1+\exp(\b\w)]$ is the Fermi function and 
$f^{>}(\w)=f^{<}(\w)-1$. 
Taking $z$ and $z'$ on the real axis but on 
different branches of the Keldysh contour, we can extract the 
lesser and greater component of the NEGF.
We first notice that for $z=t_{\pm}$ the contour evolution operators 
reduce to the standard evolution operators, i.e., 
$\ubS^{R}(t_{\pm})=\ubS(t)$ and $\ubS^{L}(t_{\pm})=\ubS^{\dag}(t)$ 
with
\begin{equation}
i\frac{d}{d t}\ubS(t)=\ubH(t)\ubS(t),\quad\quad
\ubS(0)=\uno\, ,
\label{rtmsc}
\end{equation}
and $\ubH(t)=\ubH_{\m}(t_{\pm})$, see Eq. (\ref{hmateler}). Then, in 
terms of the evolved states $\Q_{q}(t)=\ubS(t)\Q_{q}$ with components
$\Q_{q}(t)=[u_{q}(t),v_{q}(t)]$ we find
\begin{eqnarray}
\ubG^{\lessgtr}(t;t')
\equiv\ubG(t_{\mp};t'_{\pm})\!=\!
\left[
\begin{array}{cc}
    \bG_{\ua}^{\lessgtr}(t;t') & -\bF^{\gtrless,T}(t';t) \\ \\
    \overline{\bF}^{\lessgtr}(t;t') & -\bG_{\da}^{\gtrless,T}(t';t)
\end{array}
\right]
\nonumber \\
=i\sum_{q}f^{\lessgtr}(E_{q})\left[
\begin{array}{cc}
    u_{q}(t) u^{\dag}_{q}(t') & u_{q}(t) v^{\dag}_{q}(t') \\ \\
    v_{q}(t) u^{\dag}_{q}(t') & v_{q}(t) v^{\dag}_{q}(t')
\end{array}
\right]\!,
\label{ngg<>}
\end{eqnarray}
where the superscript $T$ in $\bF^{\gtrless,T}$ and 
$\bG_{\da}^{\gtrless,T}$ denotes the transpose of the matrix, see 
also Eq. (\ref{kgfcomp}).
The functions $u_{q}(t)$ and $v_{q}(t)$ can be determined by solving
a coupled system of first-order differential equations. From Eq. 
(\ref{rtmsc}) it follows that
\be
i\frac{d}{dt}u_{q}(m,t)=\sum_{n}T_{mn}e^{i\g_{mn}(t)}u_{q}(n,t)+\D_{m}(t)v_{q}(m,t),
\label{tdes1}
\ee
\be
i\frac{d}{dt}v_{q}(m,t)=-\sum_{n}T_{nm}e^{i\g_{nm}(t)}v_{q}(n,t)
+\D^{\ast}_{m}(t)u_{q}(m,t),
\label{tdes2}
\ee
which have the structure of the TD BdG 
equations.\cite{k.1969,kummelreview} As in the 
static case, however, the solution of Eqs. (\ref{tdes1}-\ref{tdes2}) 
yields the exact densities and not their BCS approximation.

We notice that for the KS system to reproduce the 
{\em time-independent} densities of an interacting system in 
equilibrium it must be 
\be
\D_{m}(t)=e^{-2i\m t}\D_{m}
\ee
for which one finds the solutions $u_{q}(t)=e^{-i(E_{q}+\m)t}u_{q}$ 
and $v_{q}(t)=e^{-i(E_{q}-\m)t}v_{q}$. The above time-dependence of 
the pairing field is 
the same as in the BCS approximation.

Using Eq. (\ref{ngg<>}) the retarded (R) and advanced (A)  
NEGF are
\bea
\ubG^{\rm R/A}(t;t')&\equiv&
\pm \theta(\pm t \mp t')
\left[\ubG^{>}(t;t')-\ubG^{<}(t;t')\right]
\nonumber \\
&=&\mp i\theta(\pm t\mp t')\ubS(t)\ubS^{\dag}(t'),
\label{gret}
\eea
with components
\be
\ubG^{\rm R/A}_{mn}(t;t')
=\left[
\begin{array}{cc}
    \bG_{\ua,mn}^{\rm R/A}(t;t') & -\bF_{nm}^{\rm A/R}(t';t) \\ \\
    \overline{\bF}_{mn}^{\rm R/A}(t;t') & -\bG_{\da,nm}^{\rm A/R}(t';t)
\end{array}
\right].
\label{gretcomp}
\ee
It follows that $\ubG^{\lessgtr}(t;t')$ can also be written as
\begin{equation}
\ubG^{\lessgtr}(t;t')=
\ubG^{\rm R}(t;0)\,\ubG^{\lessgtr}(0;0)\,
\ubG^{\rm A}(0;t').
\end{equation}

\subsection{Application to quantum transport}
\label{qtappl}

We here apply the above formalism to systems described by 
$\a=1,\ldots,{\cal N}$ bulk superconducting leads in contact 
with a central region $C$ which can be, e.g., a quantum dot, a 
molecule or a nanostructure. Assuming no direct coupling between 
the leads the Hamiltonian $\ubH_{\m}$ is written in terms of its 
projections on different subspaces as
\be
\ubH_{\m}=\sum_{\a=1}^{\cal N}\ubH_{\m,\a\a}+\ubH_{\m,CC}+
\sum_{\a=1}^{\cal N}(\ubH_{\m,\a C}+\ubH_{\m,C\a}),
\ee
where $\ubH_{\m,\a\a}$ describes the $\a$-th lead, 
$\ubH_{\m,CC}$ the nanostructure $C$ and $\ubH_{\m,\a C}+\ubH_{\m,C\a}$ 
the coupling between lead $\a$ and $C$.
We assume region $C$ to be a constriction so small that 
the bulk equilibrium of the leads is not altered by the coupling to 
$C$.
Furthermore we consider time-dependent perturbations 
which correspond to the switching on of a longitudinal electric field in 
lead $\a$. 
The time to screen the external electric field in the leads 
is in the plasmon time-scale region.
If we are interested in external fields which vary on a 
much longer time-scale it is reasonable to expect that the leads 
remain in local equilibrium. 
Therefore the coarse-grained time evolution 
of the system can be described by the 
following  TD Hamiltonian $\ubH_{\m}(t_{\pm})=\ubH(t)$
\be
\ubH_{\a\a}(t)=\exp\left(-i\m t \s_{z}\right)
\ubH_{\a\a}(0)\exp\left(i\m t \s_{z}\right),
\label{haat}
\ee
\be
\ubH_{\a C}(t)=\exp\left(i\int_{0}^{t}d\bar{t}\,
U_{\a}(\bar t)\s_{z}\right)\ubH_{\a C}(0),
\label{hact}
\ee
\be
\ubH_{C\a}(t)=[\ubH_{C\a}(t)]^{\dag}.
\label{hcat}
\ee
We do not specify the time dependence of $\ubH_{CC}(t)$ since it can 
be any, see below.
The TD field $U_{\a}(t)$ is the sum of the external and 
Hartree field and is homogeneous, i.e., it does not 
carry any dependence on the internal structure of the leads, 
in accordance with the above discussion. It has been shown that for macroscopic leads
the assumption of homogeneity is verified with rather high accuracy.\cite{mssvl.2009}

As for the case of normal leads the equations of motion for the 
Keldysh-Green's function can be solved by an embedding procedure. We 
define the uncontacted Green's function $\ubg$ which obeys 
the equations of motion (\ref{lem},\ref{rem}) with $\ubH_{\m,\a 
C}=\ubH_{\m,C\a}=0$  and 
the same boundary conditions as $\ubG$. Then, 
the equation of motion for 
$\ubG_{CC}$ projected onto regions $CC$ takes the form
\bea
\left\{i\frac{\overrightarrow{d}}{d z}\,\uno_{CC}
-\ubH_{\m,CC}(z)\!\right\}\!\ubG_{CC}(z;z')=\uno_{CC}\d(z-z')
\nonumber \\
+\!\!\int d
\bar{z}\,\ubgS(z;\bar{z})\,\ubG_{CC}(\bar{z};z'),
\label{lemcc}
\eea
where the embedding self-energy is expressed in terms of $\ubg$ as
\bea
\ubgS(\bar{z};\bar{z}')&=&\sum_{\a=1}^{\cal N}
\ubgS_{\a}(\bar{z};\bar{z}')    
\nonumber \\
&=&\sum_{\a=1}^{\cal N}\ubH_{\m,C\a}(\bar{z})\,
\ubg_{\a\a}(\bar{z};\bar{z}')\,\ubH_{\m,\a C}(\bar{z}').
\label{segen}
\eea
The above equation of motion is defined on the  
Keldysh contour of Fig. \ref{keldyshc}. Converting Eq. (\ref{lemcc}) in equations for 
real times results in a set of coupled equations known as 
Kadanoff-Baym equations\cite{book2,d.1984,kb.2000,dvl.2007,sdvl.2009,vfva.2009} recently implemented to 
study transient responses of interacting electrons in model molecular 
junctions.\cite{mssvl.2008,mssvl.2009}
The use of the Kadanoff-Baym equations to address transient and
relaxation effects in other contexts has been pioneered by 
Sch\"afer,\cite{s.1996} Bonitz et al.,\cite{bksbkk.1996} and Binder et 
al..\cite{bkbk.1997}

The importance of using an uncontacted Green's function $\ubg$ with 
boundary conditions (\ref{bcsc1},\ref{bcsc2}) for a proper 
description of $\ubG^{\lessgtr}(t;t')$ at finite times
has been discussed elsewhere in the context of transient 
regimes\cite{sa.2004,mssvl.2009} and it has been shown that it leads 
to coupled equations between the Keldysh-Green's function with two 
real times and those with one real and one imaginary time. 

In the next Section we propose a wave-function based propagation scheme 
to solve Eq. (\ref{lemcc}) for TD Hamiltonians of the form 
(\ref{haat}-\ref{hcat}).

\section{Numerical Algorithm}
\label{na}

We consider semi-infinite periodic leads with a supercell of 
dimension $N_{\rm cell}^{\a}$ for lead $\a$.  
The projected Hamiltonian $\ubH_{0,\a\a}=\ubH_{\a\a}(0)$ can then  be organized as follows
\begin{equation}
\ubH_{0,\a\a}=\left[
\begin{array}{cccc}
    \ubh_{\a} & \ubt_{\a} & \zero_{\a} & \ldots \\
    \ubt^{\dag}_{\a} & \ubh_{\a} & \ubt_{\a} & \ldots \\ 
    \zero_{\a} & \ubt^{\dag}_{\a} & \ubh_{\a} & \ldots \\ 
    \ldots & \ldots & \ldots & \ldots
\end{array}    
\right],
\label{trdiag}
\ee
where $\ubh_{\a}$ is the $2N_{\rm cell}^{\a}\times 2 N_{\rm cell}^{\a}$
Nambu Hamiltonian of the supercell with matrix structure
\be
\ubh_{\a}=\left[
\begin{array}{cc}
\bge_{\a} & \bgD_{\a} \\
\bgD^{\ast}_{\a} & -\bge_{\a}^{T}
\end{array}
\right],
\ee
while $\ubt_{\a}$ describes the contact between two nearest neighbor 
supercells. Since the pairing field is local the off-diagonal terms 
of $\ubt_{\a}$ are zero and therefore the general structure of the 
hopping matrix is
\be
\ubt_{\a}=\left[
\begin{array}{cc}
\bt_{\a} & \ze_{\a} \\
\ze_{\a} & -\bt_{\a}^{T}
\end{array}
\right].
\end{equation}
The matrices $\bge_{\a}$, $\bgD_{\a}$ and $\bt_{\a}$  in $\ubh_{\a}$ and $\ubt_{\a}$ have the dimension of the 
unit cell, i.e., $N_{\rm cell}^{\a}\times N_{\rm cell}^{\a}$. In 
particular $\bgD_{\a}$ is a diagonal matrix.

\subsection{Calculation of initial states}
\label{initeign}

Given the above structure of the leads Hamiltonian the eigenstates of 
$\ubH_{0}-\m\us$ can be grouped in scattering states with incoming 
bogolons from lead $\a=1,\ldots,{\cal N}$ and Andreev bound states 
(ABS). 

\subsubsection{Scattering states}

The lead $\a$ is characterized by 
energy bands $E_{\n}^{\a}(p)$ with 
$\n=1,\ldots,2N_{\rm cell}^{\a}$ and $p\in(0,\p)$. For a given $p$
the energies $E_{\n}^{\a}(p)$ are the solutions of the eigenvalue problem
\be
\left(\ubh_{\a}+\ubt_{\a}e^{ip}+\ubt_{\a}^{\dag}e^{-ip}-\m \us_{\a}\right)
U_{\n p}^{\a}=E_{\n}^{\a}(p) U_{\n p}^{\a}
\ee
with $U_{\n p}^{\a}$ the Nambu-Bloch eigenvectors. 
We write the index of the localized orbital $\vf_{m}$ as  
$m=s,j,\a$;
here $s$ labels the orbital within the supercell, $j$ the supercell 
and $\a$ the lead. The index $s$ runs between 1 and $N_{\rm 
cell}^{\a}$ while the supercell index $j=0,\ldots,\inf$. The
scattering state 
for an incoming bogolon from lead $\a$ has the general form
\begin{widetext}
\begin{equation}
\Q^{\a}_{\n p}(m)=\left\{
\begin{array}{ll}
U^{\a}_{\n p}(s)e^{-ipj}+
\sum\limits_{\r}R^{\a\a}_{\n p,\r}\,
W^{\a\a}_{\n p,\r}(s)\,e^{iq^{\a\a}_{\n p,\r}j} & \quad m=s,j,\a\\
& \\
\Q^{\a}_{\n p,C}(m) & \quad  m\in C \\
&  \\
\sum\limits_{\r}
T^{\a\b}_{\n p,\r}\,W^{\a\b}_{\n p,\r}(s)
e^{iq^{\a\b}_{\n p,\r}j} & 
\quad m=s,j,\b\neq \a
\end{array}
\right.
\label{scattstates}
\end{equation}
\end{widetext}
with reflection coefficients $R$ and transmission coefficients $T$.
The momenta $q^{\a\b}_{\n p,\r}$ (for all leads $\b$ including 
$\b=\a$) are associated to states with energy 
$E=E_{\n}^{\a}(p)$ and can therefore be obtained from the roots of 
\be
{\rm 
Det}[\ubh_{\b}+\ubt_{\b}e^{iq}+\ubt_{\b}^{\dag}e^{-iq}-\m\us_{\b}-E\uno_{\b}]=0.
\ee
The above equation admits, in general, complex solutions for $q$. In Eq. (\ref{scattstates})
the sums over $\r$ run over real solutions $q$ for which the sign of the Fermi 
velocity $v^{\b}_{\r}(q)=\de E^{\b}_{\r}(q)/\de q$ is opposite to the 
sign of the Fermi velocity $v^{\a}_{\n}(p)$ of the incoming bogolon 
and over all complex solutions $q$ for which ${\rm Im}[q]>0$ 
(evanescent states). Once the 
$q^{\a\b}_{\n p,\r}$ are known the Bloch state  
$W^{\a\b}_{\n p,\r}$ is simply the eigenvector with zero eigenvalue of 
the matrix $\ubh_{\b}+\ubt_{\b}e^{iq^{\a\b}_{\n p,\r}}+
\ubt_{\b}^{\dag}e^{-iq^{\a\b}_{\n p,\r}}-\m\us_{\b}-E\uno_{\b}$.
For the calculation of the reflection and transmission coefficients 
as well as of the amplitude $\Q^{\a}_{\n p,C}(m)$ in the central 
region we extended a recently proposed wave-guide approach.\cite{spbc.2009}
The method is based on projecting the Schr\"odinger equation 
$(\ubH_{0}-\m\us)\Q=E\Q$ onto the central region and onto all the 
supercells in contact with the central region, i.e., with $j=0$. 
The projection onto a $j=0$ supercell leads to an equation which 
couples the amplitude of $\Q$ in $j=0$ with that in $j=1$.
Exploiting the analytic form of the eigenstate in Eq. (\ref{scattstates}) 
the amplitude in the leads can entirely be expressed in terms of the unknown $R$'s and 
$T$'s for all $j$. In this way the equations can be closed and the 
problem is mapped into a simple linear system of equations for the unknown 
$R^{\a\b}_{\n p,\r}$, $T^{\a\b}_{\n p,\r}$ and $\Q^{\a}_{\n p,C}(m)$.

\subsubsection{Andreev bound states}
\label{abssec}

The presence of a gap in the spectrum of the superconducting leads may lead to 
the formation of localized ABS within the gap.
The procedure to calculate the ABS is slightly different 
from the one previously presented since the ABS energy is not an input 
parameter and the ABS state is normalized to 1 over the whole system.
The energy $E_{b}$ of an ABS $\Q_{b}$ is outside the lead continua. 
Projecting the Schr\"odinger equation 
$(\ubH_{0}-\m\us)\Q_{b}=E_{b}\Q_{b}$ onto different regions and 
solving for the projection $\Q_{b,C}$ in region $C$ one finds 
$(\ubH^{\rm eff}_{0,CC}(E_{b})-\m\us_{CC})\Q_{b,C}=E_{b}\Q_{b,C}$ where
\be
\ubH^{\rm eff}_{0,CC}(E)\!=\!\ubH_{0,CC}
+\!\sum_{\a}\ubH_{0,C\a}
\frac{1}{E\!-\!(\ubH_{0,\a\a}\!\!-\!\m\us_{\a\a})}
\ubH_{0,\a C}.
\ee
The ABS energies $E_{b}$ can then be extracted from the roots 
of ${\rm Det}[\ubH^{\rm eff}_{0,CC}(E)-\m\us_{CC}-E\uno_{CC}]=0$ and the 
eigenvector with zero eigenvalue of 
$\ubH^{\rm eff}_{0,CC}(E_{b})-\m\us_{CC}-E_{b}\uno_{CC}$ is proportional to the 
projection $\Q_{b,C}$ of the ABS in region $C$. We call $C_{b}$ the 
unknown constant of proportionality.
As for the scattering states we can construct the ABS everywhere 
in the system according to
\begin{equation}
\Q_{b}(m)=\left\{
\begin{array}{ll}
\sum\limits_{\r}B^{\a}_{b,\r}W^{\a}_{b,\r}(s)e^{iq^{\a}_{b,\r}j} &
\quad m=s,j,\a \\ & \\
\Q_{b,C}(m) & \quad m\in C
\end{array}
\right..
\label{absstates}
\end{equation}
The momenta $q^{\a}_{b,\r}$ and Bloch states $W^{\a}_{b,\r}$ are calculated in 
the same way as for the scattering states.  By definition all momenta
have a finite imaginary part and the sum in Eq. (\ref{absstates}) runs 
over those with a positive imaginary part. The constants 
$B^{\a}_{b,\r}$ can be simply obtained by projecting the Schr\"odinger equation 
$(\ubH_{0}-\m\us)\Q_{b}=E_{b}\Q_{b}$ onto the supercells in contact 
with region $C$, i.e., with $j=0$. The resulting equation couples the 
amplitude of $\Q_{b}$ in $j=0$ with that in $j=1$ and with the known 
amplitude $C_{b}\Q_{b,C}(m)$. Exploiting the analytic form of 
$\Q_{b}$ in the leads the amplitude in $j=1$ can entirely be 
expressed in terms of the constants $C_{b}B^{\a}_{b,\r}$ thus 
yielding a linear system of equations for each lead. Once the 
$C_{b}B^{\a}_{b,\r}$ are known the constant of proportionality 
$C_{b}$ is 
fixed by imposing that the ABS is normalized to 1. This can be easily 
done since the sums over $j$ are geometrical series.

\subsection{Embedded Crank-Nicholson propagation scheme}

To propagate the generic eigenstate $\Q$ of $\ubH_{0}-\m\us$ we 
extend the embedded Crank-Nicholson\cite{ksarg.2005,spc.2008} scheme to superconducting 
leads. The equations of motion (\ref{tdes1},\ref{tdes2}) can be 
written in a compact form as 
\be
i\frac{d}{dt}\Q(t)=\ubH(t)\Q(t), \quad \Q(0)=\Q
\label{eomgen}
\ee
where the components of the TD Hamiltonian are given in Eqs. 
(\ref{haat}-\ref{hcat}). We first perform the gauge transformation
$\Q_{\a}(t)=\exp[-i\m\us_{\a\a} t]\F_{\a}(t)$ for the projection of the 
state $\Q$ onto lead $\a$ and $\Q_{C}(t)=\F_{C}(t)$ for region 
$C$. The state $\F(t)$ obeys the equation
\be
i\frac{d}{dt}\F(t)=\tilde{\ubH}(t)\F(t), \quad \F(0)=\Q
\label{eomgengt}
\ee
with 
\be
\tilde{\ubH}_{\a\a}(t)=\ubH_{\a\a}(0)-\m\us_{\a\a}
\label{tildeaa}
\ee
\be
\tilde{\ubH}_{\a C}(t)=\exp\left[i\left(\m t+\int_{0}^{t}d\bar{t}
\,U_{\a}(\bar{t})\right)\us_{\a\a}\right]\ubH_{\a C}(0)
\label{tdcont}
\ee
and $\tilde{\ubH}_{CC}(t)=\ubH_{CC}(t)$. The advantage of the gauge 
transformed equations is that the lead Hamiltonian is now 
independent of time. We discretize the time as $t_{m}=2m\d$ 
and define $\F^{(m)}=\F(t_{m})$ and 
$\tilde{\ubH}^{(m)}=\frac{1}{2}\left[
\tilde{\ubH}(t_{m+1})+\tilde{\ubH}(t_{m})\right]$. The differential 
operator in Eq. (\ref{eomgengt}) is then approximated by the
Cayley propagator
\begin{equation}
\left(
\uno+i\d\tilde{\ubH}^{(m)}\right)
\F^{(m+1)}=
\left(
\uno-i\d\tilde{\ubH}^{(m)}\right)
\F^{(m)}.
\label{cnsc}
\end{equation}
The above propagation scheme is known as Crank-Nicholson algorithm 
and it is norm-conserving and accurate up to second order in $\d$. As 
the matrix $\tilde{\ubH}$ is infinite dimensional the direct implementation of
Eq. (\ref{cnsc}) is not possible. A significant progress can be done 
using an embedding procedure which, as we shall see, entails 
perfect transparent boundary conditions at the interfaces between 
region $C$ and leads $\a$.
Projecting Eq. (\ref{cnsc}) onto lead $\a$ and iterating one 
finds
\bea
\F_{\a}^{(m+1)}\!=
\ubg_{\a\a}^{m+1}\F_{\a}^{(0)}\!
-\frac{i\d}{\uno_{\a\a}+i\d\tilde{\ubH}_{\a\a}}\sum_{j=0}^{m}\,
\ubg_{\a\a}^{j}\tilde{\ubH}_{\a C}^{(m-j)}
\nonumber \\ 
\times\left(
\F_{C}^{(m+1-j)}+\F_{C}^{(m-j)}\right),
\label{pham+1}
\eea
where we have defined the propagator
\begin{equation}
\ubg_{\a\a}=
\frac{\uno_{\a\a}-i\d\tilde{\ubH}_{\a\a}}{\uno_{\a\a}+i\d\tilde{\ubH}_{\a\a}},
\end{equation}
and made use of the fact that 
$\tilde{\ubH}_{\a\a}(t)\equiv\tilde{\ubH}_{\a\a}$ is time-independent.
The time-dependence of the contacting Hamiltonian can be easily 
extracted from Eq. (\ref{tdcont}) and reads
\begin{equation}
\tilde{\ubH}_{\a C}^{(m)}=
\frac{\exp\left(i\m_{\a}^{(m+1)}\us_{\a\a}\right)+
\exp\left(i\m_{\a}^{(m)}\us_{\a\a}\right)}{2}
\tilde{\ubH}_{\a C}(0),
\label{chrsc}
\end{equation}
where we have defined 
\be
\m_{\a}^{(m)}=\m t_{m}+\int_{0}^{t_{m}}d\bar{t}\,U_{\a}(\bar{t}).
\ee
At this point comes a crucial observation which allows for extending the 
propagation scheme  of Refs. \onlinecite{ksarg.2005,spc.2008} to the 
superconducting case. Since the pairing field is local in the chosen 
basis the off-diagonal part of the contacting Hamiltonian is zero and 
hence $\tilde{\ubH}_{C\a}\us_{\a\a}=\us_{CC}\tilde{\ubH}_{C\a}$. It 
follows that Eq. (\ref{chrsc}) can also be rewritten as
\bea
\tilde{\ubH}_{\a C}^{(m)}&=&
\tilde{\ubH}_{\a C}(0)
\frac{\exp\left(i\m_{\a}^{(m+1)}\us_{CC}\right)+
\exp\left(i\m_{\a}^{(m)}\us_{CC}\right)}{2}
\nonumber \\
&\equiv&
\tilde{\ubH}_{\a C}(0)\bar{\ubcalz}_{\a}^{(m)},
\label{chrsc2}
\eea
which implicitly define the matrices 
$\bar{\ubcalz}_{\a}^{(m)}=(\ubcalz_{\a}^{(m)})^{\ast}$.
Next we project Eq. (\ref{cnsc}) onto region $C$ and use Eq. 
(\ref{pham+1}) to express the $\F_{\a}$ at a given time step in terms of 
the $\F_{C}$ at all previous time steps. The resulting equation is
\begin{eqnarray}
\left(
\uno_{CC}+i\d\tilde{\ubH}_{\rm eff}^{(m)}
\right)
\F_{C}^{(m+1)}&=&
\left(
\uno_{CC}-i\d\tilde{\ubH}_{\rm eff}^{(m)}
\right)
\F_{C}^{(m)}
\nonumber \\
&+&\sum_{\a}\left(S_{\a}^{(m)}+M_{\a}^{(m)}\right)
\label{maintd}
\eea
and contains only quantities with the dimension of region $C$.
We emphasize that Eq. (\ref{maintd}) is an exact reformulation of the 
original Eq. (\ref{cnsc}) but it has the advantage of being implementable. 
Indeed, exploiting the result in Eq. (\ref{chrsc2}) 
the boundary term $S_{\a}^{(m)}$ and memory term $M_{\a}^{(m)}$ read
\be
S_{\a}^{(m)}=-i\d
\ubcalz_{\a}^{(m)}
\tilde{\ubH}_{C\a}(0)\,
\ubg_{\a\a}^{m}\left(\uno_{\a\a}+\ubg_{\a\a}\right)
\F_{\a}^{(0)},
\label{bndrterm}
\ee
\bea
M_{\a}^{(m)}=
-\d^{2}\sum_{j=0}^{m-1}
\ubcalz_{\a}^{(m)}
\left(\ubQ_{\a}^{(j+1)}+
\ubQ_{\a}^{(j)}
\right)\bar{\ubcalz}_{\a}^{(m-1-j)}
\nonumber \\
\times
\left(
\F_{C}^{(m-j)}+\F_{C}^{(m-1-j)}\right),
\label{memterm}
\eea
while the effective Hamiltonian is given by
\begin{equation}
\tilde{\ubH}_{\rm eff}^{(m)}=
\tilde{\ubH}_{CC}^{(m)}-i\d\sum_{\a}
\ubcalz_{\a}^{(m)}\ubQ_{\a}^{(0)}\bar{\ubcalz}_{\a}^{(m)},
\label{heffscnw}
\end{equation}
where the embedding matrices $\ubQ_{\a}^{(m)}$ have twice the 
dimension of region $C$ and are defined according to
\begin{equation}
\ubQ_{\a}^{(m)}=
\tilde{\ubH}_{C\a}(0)\frac{\left(\uno_{\a\a}-i\d\tilde{\ubH}_{\a\a}\right)^{m}}
{\left(\uno_{\a\a}+i\d\tilde{\ubH}_{\a\a}\right)^{m+1}}\,\tilde{\ubH}_{\a C}(0).
\label{qcffsc}
\end{equation}
In Appendix \ref{embq} we describe a recursive scheme to calculate 
the embedding matrices. In Appendix \ref{bndtermcalc} we further show 
that the boundary term $S_{\a}^{(m)}$ can be expressed in terms of the $\ubQ_{\a}$'s 
thus rendering Eq. (\ref{maintd}) a well defined equation for time 
propagations.

In the next Section we apply the numerical scheme to UF-JNJ model systems 
and obtain results for the TD densities and currents.

\section{Real-time simulations of S-D-S junctions}
\label{rts}

Due to the vast phenomenology of S-D-S junctions it is not possible to 
address these systems in a single work. Furthermore the analysis of 
the time-dependent regime is generally more complex than that in 
the Josephson regime and  it is therefore advisable to first gain some insight 
by investigating simple cases.
Our intention in this Section 
is to demonstrate the feasibility of the propagation scheme and to 
present genuine TD properties of simple model systems.

We consider a tight-binding chain (region $C$) with nearest neighbor hopping 
$t_{C}$ and on-site energy $\e_{C}$ connected to a left ($L$) and 
right ($R$) wide-band leads.
The $\a=L,R$ lead is described by a semi-infinite tight-binding chain 
with nearest neighbor hopping $t_{\a}$ and a constant pairing 
field $\D_{\a}$, and is coupled to the $\a$ end-point of the 
central chain through its surface site with a hopping $t_{C\a}=t_{\a C}$. The 
system is initially in equilibrium at temperature $T=0$ and chemical 
potential $\mu=0$ and driven out of equilibrium by a TD  bias voltage $U_{\a}(t)$ 
applied to lead $\a$ at positive times. From Section \ref{qtappl},
the Hamiltonian for this kind 
of systems read
$\hat{H}(t)=\sum_{\a}(\hat{H}_{\a\a}(t)+
\hat{H}_{\a C}(t)+\hat{H}_{C\a}(t))+\hat{H}_{CC}$ where 
\bea
\hat{H}_{\a\a}(t)&=&t_{\a}\sum_{j=0}^{\inf}
\sum_{\s}(\hat{c}^{\dag}_{j+1\s\a}\hat{c}_{j\s\a}+{\rm H.c.})
\nonumber \\
&+&(e^{-2i\m t}\D_{\a}\hat{c}^{\dag}_{j\ua\a}\hat{c}^{\dag}_{j\da\a}
+{\rm H.c.})
\eea
describes the lead $\a=L,R$, 
\be
\hat{H}_{L C}(t)=t_{LC}e^{i\int_{0}^{t}dt'U_{L}(t')}\sum_{\s}
\hat{c}^{\dag}_{0\s L}\hat{c}_{0\s}+{\rm H.c.}
\ee
\be
\hat{H}_{R C}(t)=t_{RC}e^{i\int_{0}^{t}dt'U_{R}(t')}\sum_{\s}
\hat{c}^{\dag}_{0\s R}\hat{c}_{N\s}+{\rm H.c.}
\ee
accounts for the coupling between region $C$ and the leads, and
\be
\hat{H}_{CC}=t_{C}\sum_{m=0}^{N-1}\sum_{\s}
(\hat{c}^{\dag}_{m+1\s}\hat{c}_{m\s}+{\rm H.c.})
+\e_{C}\sum_{m=0}^{N}\sum_{\s}\hat{c}^{\dag}_{m\s}\hat{c}_{m\s}
\label{hccsp}
\ee
is the Hamiltonian of the chain with $N+1$ atomic sites.
The currents $J_{L}(t)\equiv J_{0L,0}(t)$ and $J_{R}(t)\equiv J_{N,0R}(t)$ 
through the bonds connecting the chain to the left and right leads 
are obtained from Eq. (\ref{avecurr}) and Eq. (\ref{ngg<>}) and 
read
\bea
J_{L}(t)&=&-it_{LC}e^{i\g_{LC}(t)}\left[\sum_{q}
f^{<}(E_{q})u_{q}(0L,t)u^{\ast}_{q}(0,t)
\right.
\nonumber \\ 
&-&\left.\sum_{q}
f^{>}(E_{q})v_{q}(0,t)v^{\ast}_{q}(0L,t)
\right]
+{\rm H.c.},
\label{jjl}
\eea
\bea
J_{R}(t)&=&-it^{\ast}_{RC}e^{-i\g_{RC}(t)}\left[\sum_{q}
f^{<}(E_{q})u_{q}(0,t)u^{\ast}_{q}(0R,t)
\right.
\nonumber \\ 
&-&\left.\sum_{q}
f^{>}(E_{q})v_{q}(0R,t)v^{\ast}_{q}(0,t)
\right]
+{\rm H.c.},
\label{jjr}
\eea
where $\g_{\a C}(t)=i\int_{0}^{t}dt'U_{\a}(t')$ and
the sum over $q$ runs over all ABS and scattering states. 
Similarly, the pairing density $P_{m}(t)$ on an arbitrary site of the 
chain is obtained from Eq. (\ref{avepden}) and Eq. (\ref{ngg<>}) 
and reads
\be
P_{m}(t)=\sum_{q}f^{<}(E_{q})u_{q}(m,t)v_{q}^{\ast}(m,t)e^{2i\e_{C}t}.
\ee

We will write the pairing field as 
$\D_{\a}=\x_{\a}e^{i\chi_{\a}}\D$ and measure energies in units of $\D$, 
times in units of 
$\hbar/\D$ and currents in units of $|e|\D/\hbar$, with 
$|e|$ the absolute charge of the carriers. Since we consider
wide-band leads with $t_{\a}\gg t_{\a C},t_{C}$ and the chemical 
potential is set to zero the results depend only on the ratio 
$\G_{\a}\equiv 2t_{\a C}^{2}/t_{\a}$ (tunneling rate) and not on $t_{\a C}$ and 
$t_{\a}$ separately. In the following we therefore specify the value 
of $\G_{\a}$ only. In practical calculations the longitudinal vector 
$p\in(0,\p)$ of the scattering states, see Eq. (\ref{scattstates}), 
is discretized with $N_{p}$ mesh points and only states with energy 
within the range $(\m-\L,\m+\L)$ are propagated in time. 
We will call $N_{p,\a}$ the number of scattering states from lead $\a$
that are propagated.
The cutoff 
$\L$ is chosen about an order of magnitude larger than the typical 
energy scales of the problem, i.e., $U_{\a}$, $\G_{\a}$, $\D_{\a}$, 
$t_{C}$, $\e_{C}$. 

\subsection{The single-level quantum dot model}

The single-level quantum dot (QD) model  
corresponds to a central chain with only one atomic site ($N=0$). 
For $\D_{L}=\D_{R}=0$ (N-QD-N) the TD response of 
this system has been investigated by several 
authors
and an analytic formula for the TD current is also 
available.\cite{jmw.1994,sa.2004,szb.2009}
Scarce attention, however, has been devoted to the
system with one superconducting lead\cite{xsw.2007} (N-QD-S) and to the best of our 
knowledge the only available results when both leads are 
superconducting (S-QD-S) have been published in Ref. 
\onlinecite{psc.2009}.

\subsubsection{N-QD-S model under DC bias}

\begin{figure}[t]
\centering
\includegraphics[width=0.48\textwidth]{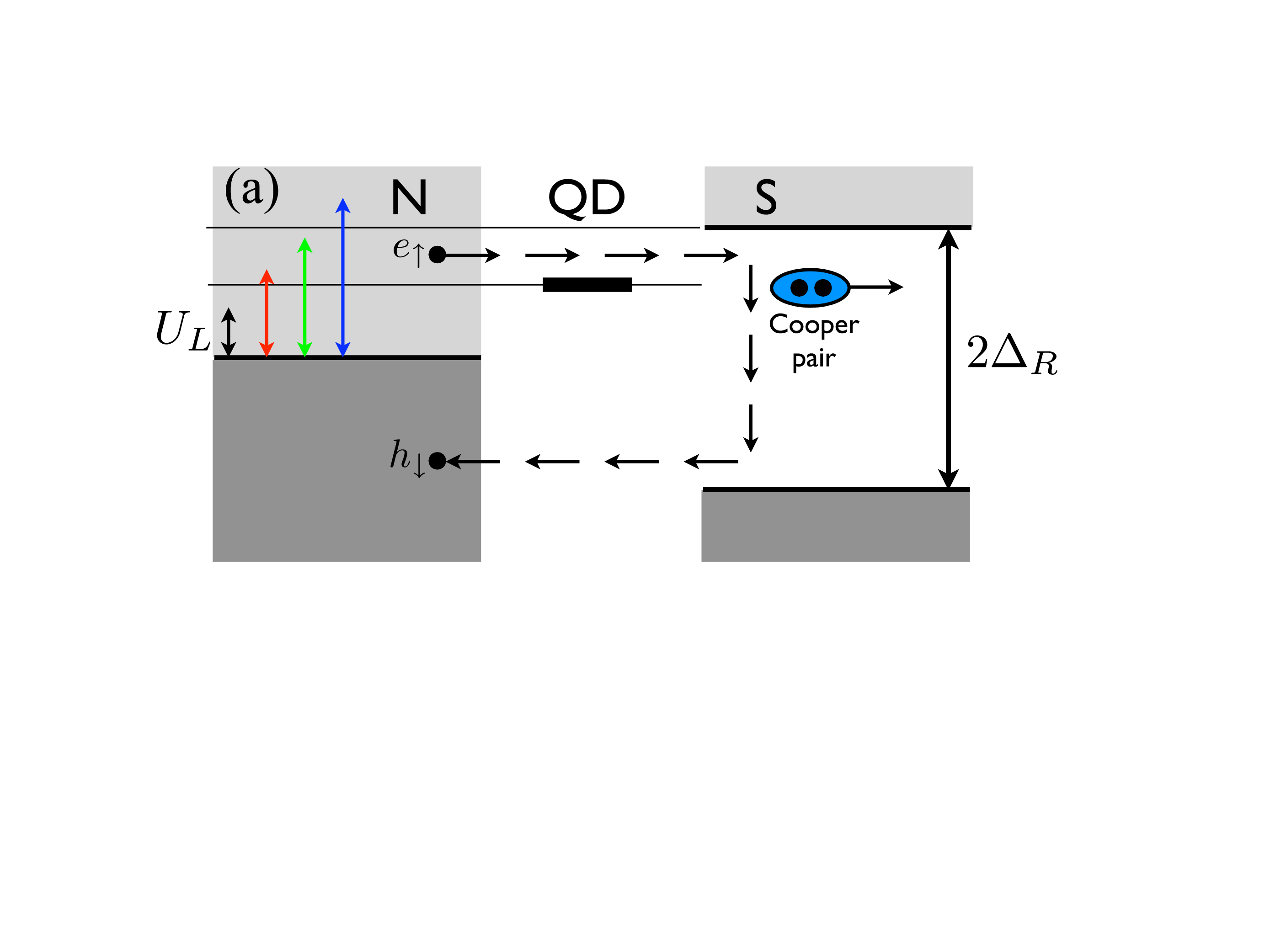}
\includegraphics[width=0.48\textwidth]{fig03.eps}
\caption{(a) Schematic of the transport set up. A single level QD 
with on-site energy $\ve_{C}=0.5$ is weakly 
connected ($\G_{L}=\G_{R}=0.2$) to a left normal lead and a right 
superconducting lead. In equilibrium both temperature $T$ and chemical 
potential $\m$ are zero. The 
system is driven out of equilibrium by a step-like voltage bias 
$U_{L}=0.3,\,0.6,\,0.9,\,1.2$ 
in the normal lead. For $U_{L}<\D_{R}$ the dominant scattering 
mechanism is the AR in which an electron is reflected as a hole and a 
Cooper pair is formed in lead $R$.
(b) Time-dependent current at the left interface 
(first panel), right interface (second panel) and absolute value of 
the pairing density on the QD (third panel). The insets show the TD 
current for the same parameters but $\D_{R}=0$, i.e., for a normal 
$R$ lead. The results are obtained with a time-step $\d=0.05$, cutoff 
$\L=6$ and a 
number of scattering states $N_{p,L}=1070$, $N_{p,R}=1056$. }
\label{fig_nqds}
\end{figure}

We first consider the N-QD-S case schematically illustrated in Fig. 
\ref{fig_nqds}(a). To highlight the different scattering mechanisms we 
shift the central level by $\e_{C}=0.5$, choose weak couplings to the 
leads $\G_{L}=\G_{R}=0.2$, and drive the system out of 
equilibrium by applying four different biases $U_{L}=0.3,\,0.6,\,0.9,\,1.2$ 
to the left normal lead. For biases in the subgap region, i.e., 
$U_{L}<\D_{R}=1$, transport is dominated by Andreev reflections (AR). 
In Fig. \ref{fig_nqds}(b) we show the currents $J_{L}(t)$ 
and $J_{R}(t)$ of Eqs. (\ref{jjl},\ref{jjr}).
For $U_{L}=0.3<\e_{C}$ the AR are strongly suppressed since 
electrons at the left electrochemical potential $\m_{L}=U_{L}$ have just 
enough energy to enter the resonant window $(\e_{C}-2\G,\e_{C}+2\G)$, 
where $2\G=\G_{L}+\G_{R}$. Resonant AR can occur for $U_{L}>\e_{C}$ 
and constitute the dominant mechanism for electron tunneling. This is 
clearly visible in the second panel of Fig. \ref{fig_nqds}(b) where 
the steady-state values of $J_{R}$ for $U_{L}=0.6$ and $U_{L}=0.9$ 
are approximatively the same. At larger biases $U_{L}=1.2>\D_{R}$ 
electrons can also tunnel via standard quasi-particle scattering and 
the steady-state current increases. This interpretation is confirmed 
by the behavior of the pairing density $P_{0}(t)$ on the QD, third 
panel of Fig. \ref{fig_nqds}(b).
For times up to $\sim 5$ the pairing density decreases since 
pre-existent Cooper pairs in lead $R$ move away from the QD. However, 
while $|P_{0}(t)|$ remains below its 
equilibrium value at $U_{L}=0.3$, for all other biases, 
$U_{L}>\e_{C}$, 
$|P_{0}(t)|$ increases after $t\sim 5$, meaning that {\em a Cooper 
pair is forming at the interface}. We also notice that the values of
$|P_{0}(t\ra\inf)|$ for $U_{L}=0.9$ and $U_{L}=1.2$ 
are very close while the corresponding currents $J_{R}$ differ appreciably. This is again 
in agreement with the fact that electrons with energy larger than 
$\D_{R}$ do not undergo AR and thus no extra Cooper pairs are formed.
Finally we observe that the transient regime is longer in the N-QD-S 
case than in the N-QD-N case, see inset in panel 2 and 3 of Fig. 
\ref{fig_nqds}(b), as also pointed out in Ref. \onlinecite{xsw.2007}.

\subsubsection{S-QD-S model under DC bias}

\begin{figure}[t]
\centering
\includegraphics[width=0.48\textwidth]{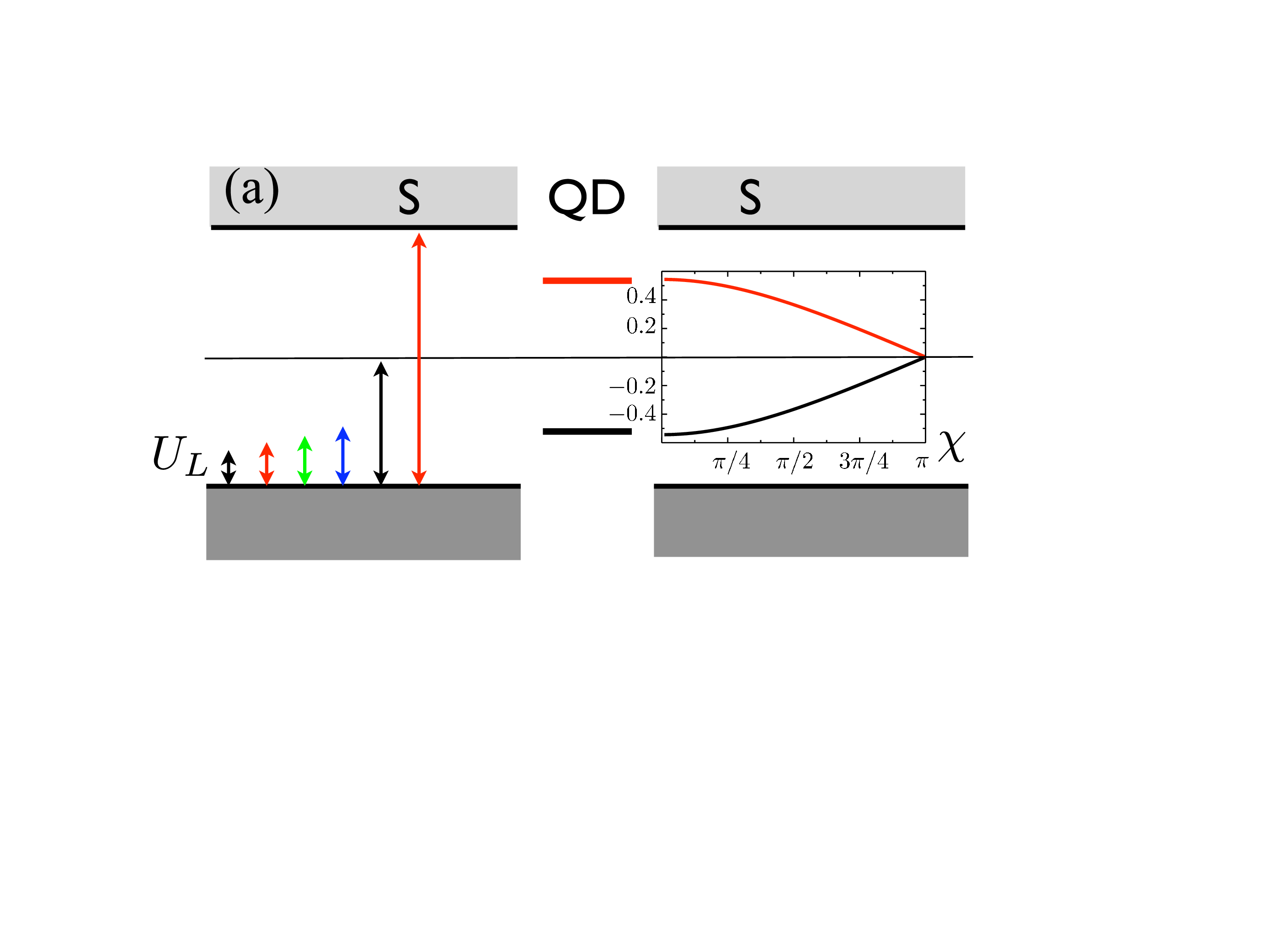}
\includegraphics[width=0.48\textwidth]{fig05.eps}
\caption{(a) Schematic of the S-QD-S model with $\G_{L}=\G_{R}=1.0$, 
$\D_{L}=\D_{R}=1$, and $\e_{C}=0$. This system admits two ABS  in the 
gap. The ABS energy depends on the superconducting phase difference 
$\chi$ as illustrated in the inset. 
(b-c) Time-dependent current $J_{L}(t)$ at the left interface as a 
function of time for (b) 
$U_{L}=3.0,\,2.0,\,1.0$ [the curves corresponding to bias $U_{L}=n.0$ 
are shifted upward by $0.3(n-1)$] and (c) $U_{L}=0.5,\,0.4,\,0.3,\,0.2$
[the curves corresponding to bias $U_{L}=0.n$ are shifted upward by $0.6(n-2)$].
The results are obtained with a time-step $\d=0.05$, cutoff $\L=12.1$, and a number of 
scattering states $N_{p,L}=N_{p,R}=768$ for panel (b) and $\d=0.05$, 
$\L=4$, $N_{p,L}=N_{p,R}=788$ for panel (c).}
\label{td_sqs_dc}
\end{figure}

We now turn to the more interesting case in which the QD is connected 
to a left and right superconducting lead (S-QD-S), see Fig. 
\ref{td_sqs_dc}(a). 
We focus on symmetric couplings $\G_{L}=\G_{R}=\G=1$ and on 
pairing fields $\D_{L}=\D_{R}e^{i\chi}= e^{i\chi}$ with the same magnitude but 
different phase. This system {\em always} 
support two Andreev bound states (ABS) in the gap. Their energy can 
be obtained analytically from the solution of 
${\rm Det}[\ubH^{\rm eff}_{0,CC}(E)-\m\us_{CC}-E\uno_{CC}]=0$ 
(see Section \ref{abssec})
which, in terms of the dimensionless variables 
$x=E/\D$, $\g=\G/\D$ and $e=(\e_{C}-\mu)/\D$, reads
\be
x^{2}(1+\frac{\g}{\sqrt{1-x^{2}}})^{2}-e^{2}-\frac{\a^{2}\g^{2}}{1-x^{2}}=0,
\label{abszero}
\ee
where $\a=\sqrt{\frac{1+\cos\chi}{2}}$ and varies in the 
range $(0,1)$. In Fig. \ref{td_sqs_dc}(a) we plot the solutions of 
Eq. (\ref{abszero}) as a function of $\chi$ for $\e_{C}=\mu=0$.
In equilibrium and at zero temperature one ABS is fully occupied and the 
other is empty. At time $t=0$ a 
constant bias $U_{L}$ is applied to the left lead. 
In Fig. \ref{td_sqs_dc}(b) we display the TD current at the left 
interface $J_{L}(t)$ for $\chi=0$ and $U_{L}=3,\,2,\,1$. After a
transient the current oscillates in time with period $T_{J} =2\p/(2U_{L})$,
as expected. For $U_{L}>2$ the S-QD-S system behaves 
similarly to a macroscopic Josephson junction with an almost pure 
monochromatic response, albeit the average value $J_{\rm dc}$ of 
the current over a period is different from zero. For $U_{L}=1<2\D$, 
i.e., in the subgap region, 
the transient regime becomes much longer and $J_{L}(t)$ deviates from 
a perfect monochromatic function. At $U_{L}=1$ the dominant 
scattering mechanism is the single AR. 

As discussed in Ref. \onlinecite{lycldmr.1997} the presence of the resonant level modifies 
substantially the $J_{\rm dc}-V$ ($V=U_{L}-U_{R}$) characteristic and 
for $\G=1$ the subharmonic gap structure is almost entirely washed out. 
However, a very rich structure is observed in the TD 
current. In Fig. \ref{td_sqs_dc}(c) we display $J_{L}(t)$ 
for biases $U_{L}=0.5,\,0.4,\,0.3,\,0.2$. The charge carriers undergo 
multiple AR (MAR) before acquiring enough energy and escaping from the QD.
The dwelling time increases with decreasing bias and the transient 
current has a highly non-trivial behavior before the Josephson 
regime sets in. From the simulations in Fig. \ref{td_sqs_dc}(c)
at  bias $U_{L}=0.2$ the propagation time $t=250$ is not sufficient 
for the development of the Josephson oscillations.
We also observe that the smaller is the bias the larger is the 
contribution of high-order harmonics, which is  
in contrast with one would naively expect from linear response theory.

\begin{figure}[t]
\centering
\includegraphics[width=0.48\textwidth]{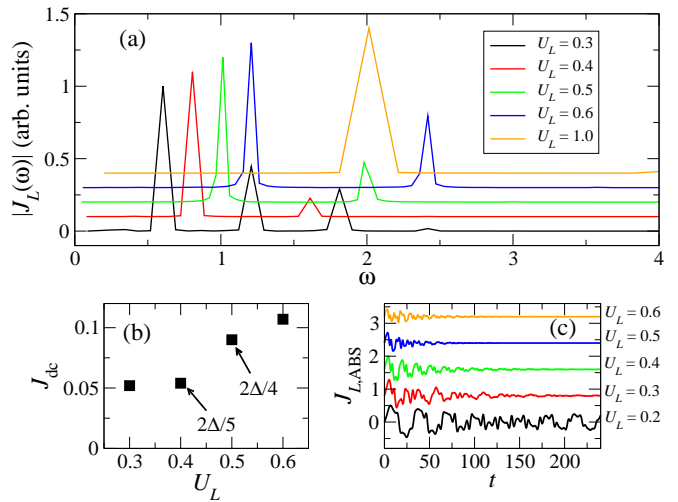}
\caption{(a) Discrete Fourier 
transform of $J_{L}(t)$ in arbitrary units [the curves corresponding to bias 
$U_{L}=0.n$ are shifted upward by $0.7(n-3)$ while that corresponding to 
bias $U_{L}=1.0$ is shifted upward by $2.8$. 
(b) Values of the average current for biases in the subgap 
region. (c) ABS contribution to the current $J_{L}(t)$ for biases 
$U_{L}=0.2,\,0.3,\,0.4,\,0.5,\,0.6$ [the curves corresponding to bias 
$U_{L}=0.n$ are shifted upward by $0.8(n-2)$].
The numerical 
parameters are the same as in Fig. \ref{td_sqs_dc}.}
\label{sqds_fourier}
\end{figure}

In Fig. \ref{sqds_fourier}(a) we display the Fourier transform of 
$J_{L}(t)-J_{\rm dc}$ in the Josephson regime. Replica of the main Josephson 
frequency $\w_{J}=2U_{L}$ are clearly visible for $U_{L}<\D$.
The values of $J_{\rm dc}$ as obtained from time propagation 
are reported in Fig. \ref{sqds_fourier}(b) and are consistent with a 
smeared sub-harmonic gap structure. 

From the curves $J_{L}(t)$ it is not evident how to estimate the 
duration of the transient time. We found useful 
to look at the contribution of the ABS, $J_{L,\rm ABS}$, 
to the total current $J_{L}$, since $J_{L,\rm ABS}(t\ra\inf)=0$. 
This quantity is evaluated from Eq. (\ref{jjl}) by restricting the sum 
over $q$ to the ABS and is 
shown in Fig. \ref{sqds_fourier}(c). ABS play a 
crucial role in the relaxation mechanism as we shall 
see in the next Section. 

\subsubsection{S-QD-S model under DC pulses}

\begin{figure}[t]
\centering
\includegraphics[width=0.48\textwidth]{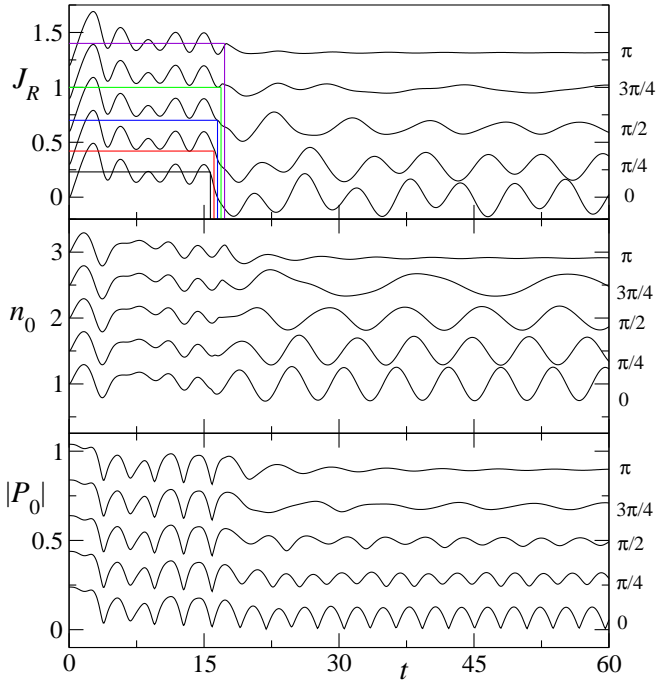}
\caption{Time-dependent current at the right interface $J_{R}$ (first 
panel) as well as the density $n_{0}$ (second panel) 
and pairing density $|P_{0}|$ (third panel) on the QD. 
The curves from bottom to top corresponds to a switch-off time 
$t^{(n)}_{\rm off}=5\p+n\p/8$, with $n=0,\,1,\,2,\,3,\,4$.
Since the bias is $U_{L}=1$ the accumulated 
phase difference $\chi^{(n)}$ at the end of the pulse 
is $\chi^{(n)}=2t^{(n)}_{\rm off}=n\p/4$. For the switch-off time 
$t^{(n)}_{\rm off}$ the curves of $J_{R}$ are shifted upward by $0.3n$,
those of $n_{0}$ by $0.5n$ and those of $|P_{0}|$ by $0.2n$. 
The results are obtained with a time-step $\d=0.05$, cutoff $\L=12.1$, and a number of 
scattering states $N_{p,L}=N_{p,R}=768$.
}
\label{stepbias}
\end{figure}

As mentioned in the introduction the possibility of employing 
UF-JNJ in future electronics rely on our understanding of 
their TD properties. In the previous Section we studied the 
transient behavior of a S-QD-S system under the sudden switch-on of 
an applied bias. Equally important is to study how the system responds 
when the bias is switched off. We therefore consider the same S-QD-S 
model as before with $\G_{L}=\G_{R}=1$, $\e_{C}=0$, 
$\D_{L}=\D_{R}=1$ initially in equilibrium at zero temperature and 
chemical potential. At time $t=0$ a constant bias $U_{L}=1$ is 
applied to lead $L$ until the time $t_{\rm off}$ at which the bias is 
switched off. How does the system relax? In Fig. \ref{stepbias} we show 
the current $J_{R}$ at the right interface as well as 
the density $n_{0}$
and pairing density $|P_{0}|$ on the QD for switch-off times 
$t^{(n)}_{\rm off}=5\p+n\p/8$ with $n=0,\,1,\,2,\,3,\,4$. Despite the 
fact that the switch-off times 
are all very close [$t^{(0)}_{\rm off}\sim 15.71$ and $t^{(4)}_{\rm 
off}\sim 17.28$] the system reacts in different ways 
and actually relaxes only in one case. 
The strong dependence on $t_{\rm off}$ is due to 
the two ABS in the gap. Similarly to what happens in normal 
systems\cite{s.2007} the asymptotic ($t\ra\inf$) form of the  
density on the QD is
\be
n_{0}(t)-n_{0,\rm cont}\sim \sum_{ij} 
f_{ij}\cos((\e^{(i)}_{\rm ABS}-\e^{(j)}_{\rm ABS})t),
\label{absdens}
\ee
where $\e^{(i)}_{\rm ABS}$, $i=1,2$, are the ABS eigenenergies of the 
Hamiltonian after the bias has been switched off 
and $n_{0,\rm cont}$ is the contribution of the 
continuum states to the density. The coefficients $f_{ij}=f_{ji}$ are matrix 
elements of the Fermi function $f(\hat{H}(0))$ calculated at the {\em 
equilibrium} Hamiltonian and depend on the 
history of the applied bias.\cite{kksg.2008,kskg.2009}
Contrary to the normal case, however, the energy of the ABS depends on 
{\em when} the bias is switched off since after a time $t_{\rm off}$ the 
phase difference $\chi$ changes from zero to $2U_{L}t_{\rm off}$.
This fact together with Eq. (\ref{absdens}) 
explains the persistent oscillations at different frequencies. Indeed  
$\chi^{(n)}=2U_{L}t^{(n)}_{\rm off}=n\p/4$ and from 
Fig. \ref{td_sqs_dc}(a) we see that 
$[\e^{(1)}_{\rm ABS}(\chi^{(n)})-\e^{(2)}_{\rm ABS}(\chi^{(n)})]$ varies 
from $\sim 1.08$ to zero when $n$ varies from zero to 4.
The amplitude of the oscillations as well as the average value of the 
density $n_{0}$, however, do not depend only on $\chi$ but also on 
the history of the applied bias. Two different biases $U_{L}(t)$ and 
$U'_{L}(t)$ yielding the same phase difference $\chi=
2\int_{0}^{t_{\rm off}}d\t U_{L}(\t)=2\int_{0}^{t_{\rm off}}d\t 
U'_{L}(\t)$ give rise to different persistent oscillations, albeit 
with the same frequency.

From the results of this Section we conclude that for devices 
coupled to superconducting leads a small difference in the switch-off 
time of the bias can cause a large difference in the relaxation time 
of the device. This property may be exploited to generate {\em zero 
bias ac currents of tunable frequency}.

\subsubsection{S-QD-S model under AC bias}

The time-propagation approach has the merit of not being 
limited to step-like biases as it can deal with any TD bias at the 
same computational cost. Of special importance is the case of ac 
biases where a microwave radiation $U_{\rm r}\sin(\w_{\rm r}t)$ is 
superimposed to a dc signal $V=U_{L}-U_{R}$. The study of UF-JNJ in the presence of 
microwave radiation started with the work of Cuevas et 
al.\cite{chmrlys.2002} who predicted the occurrence of subharmonic 
Shapiro spikes in the $J_{\rm dc}-V$ characteristic of
superconducting point contacts. Later on Zhu et al.\cite{zlml.2004} extended the 
analysis to the S-QD-S model and discuss how the ABS modify the
$J_{\rm dc}-V$ characteristic. The replicas of the Shapiro spikes 
have been experimentally observed\cite{cspjheu.2006} and
can be explained in terms of photon-assisted multiple Andreev 
reflections. Using a generalized Floquet formalism one can show that 
in the long-time limit\cite{chmrlys.2002}
\be
J_{L}(t)=\sum_{mn}J_{m}^{n}(V,\g,\w_{\rm r})e^{i(m\w_{J}+n\w_{\rm r})t}
\ee
where $\g=U_{\rm r}/\w_{\rm r}$ and $\w_{J}=2V$ is the Josephson 
frequency. The calculation of $J_{m}^{n}$ is, in general, rather 
complicated and to the best of our knowledge the full TD profile of 
$J_{L}(t)$ as well as the duration of the transient time before the 
photon-assisted Josephson regime sets in have not been addressed 
before. 

\begin{figure}[t]
\centering
\includegraphics[width=0.44\textwidth]{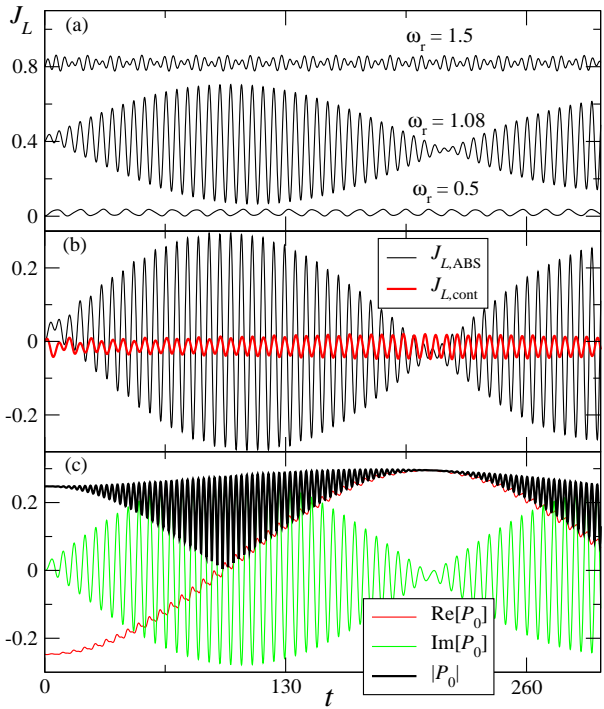}
\caption{(a) TD current at the left interface for $U_{L}=0$,  
$U_{\rm r}=0.05\w_{\rm r}$ with $\w_{\rm r}=0.5$, $1.08$ [the curve is 
shifted upward by 0.4], and $1.5$ [the curve is 
shifted upward by 0.8]. (b) ABS and continuum contribution to the 
total current in the resonant case $\w_{\rm r}=1.08$, $U_{\rm 
r}=0.05\w_{\rm r}$ and $U_{L}=0$. (c) Pairing potential on the QD for 
the same parameters as in panel (b). The results are obtained with a 
time-step $\d=0.05$, cut-off $\L=4$, and a number of scattering 
states $N_{p,L}=N_{p,R}=788$.}
\label{acfig1}
\end{figure}

\begin{figure*}[t]
\centering
\includegraphics[width=1.\textwidth]{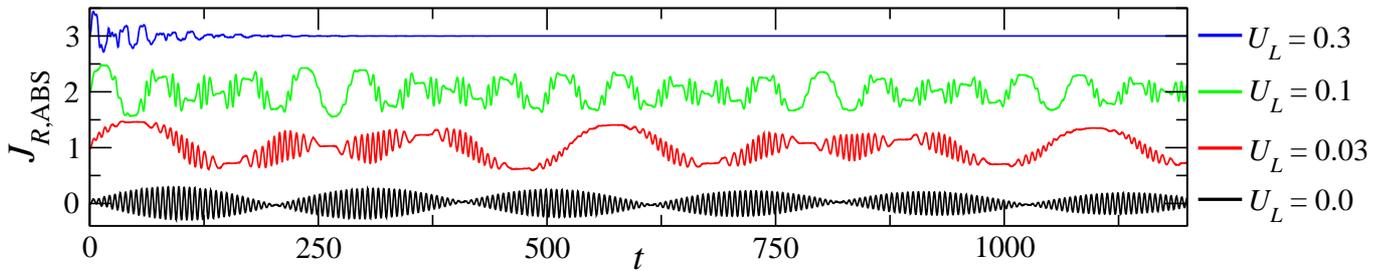}
\caption{ABS contribution to the current at the right interface for 
dc biases with a superimposed microwave radiation described by 
$U_{L}(t)=U_{L}+U_{\rm r}\sin(\w_{\rm r}t)$, with $U_{\rm 
r}=0.05\w_{\rm r}$, $\w_{\rm r}=1.08$ and
$U_{L}=0.0,\;0.03,\;0.1,\;0.3$. The system is the same as in Fig. \ref{acfig1}
with $\D_{L}=\D_{R}=1$, $\G_{L}=\G_{R}=1$ and $\e_{C}=0$. The 
time-step is $\d=0.05$.}
\label{absfig}
\end{figure*}

We here consider the S-QD-S model with $\G_{L}=\G_{R}=1$, 
$\ve_{C}=0$, $\D_{L}=\D_{R}=1$ under a dc bias and in the presence of 
a superimposed microwave radiation $U_{L}(t)=U_{L}+
U_{\rm r}\sin(\w_{\rm r}t)$ and $U_{R}=0$. In Fig. \ref{acfig1}(a) we 
display the TD current at the left interface for fixed $\g=U_{\rm 
r}/\w_{\rm r}=0.05$ and different values of the frequency $\w_{\rm 
r}=0.5,\,1.08,\,1.5$. The first striking feature is the occurrence of 
a {\em transient resonant effect} at $\w_{\rm r}=1.08\sim \w_{\rm ABS}\equiv \e^{(1)}_{\rm ABS}-\e^{(2)}_{\rm ABS}$.
At the resonant frequency the amplitude of the oscillations increases 
linearly in time till a maximum value $\sim 0.3$. The Fourier 
decomposition (not shown) reveals that the peak at $\w=1.08$ 
splits into two peaks, one above and one below 1.08, which is 
consistent with the observed beating.
The effect is absent at larger ($\w_{\rm r}=1.5$)  and smaller
($\w_{\rm r}=0.5$) frequencies for which the amplitude
of the oscillations remains below 0.05 and two main harmonics, 
one at $\w_{\rm r}$ and the other at $\w_{\rm ABS}$, 
are visible in the Fourier decomposition (not shown). 
The peak at $\w=\w_{\rm ABS}$ is due to a transient excitation with a 
long life-time and cannot be described using Floquet based 
approaches.

The ABS play a crucial role in determining the TD profile of $J_{L}$
at the resonant frequency. The total current $J_{L}(t)=J_{L,\rm 
cont}(t)+J_{L,\rm ABS}(t)$ is the sum of the current $J_{L,\rm cont}$
coming from the evolution of the continuum states and the ABS current
$J_{L,\rm ABS}(t)$. These two currents are shown in 
Fig. \ref{acfig1}(b) from which it is evident that ABS carry an 
important amount of current not only in the dc Josephson 
effect\cite{psc.2009,acz.2000} but {\em also in the transient regime}. 
In Fig. \ref{acfig1}(c) we show the pairing density on the QD for 
the resonant frequency $\w_{\rm r}=1.08$.

In the presence of an external bias the ABS contribute to the current only in the transient regime. 
The duration of the transient  is investigated in Fig. \ref{absfig} 
where we show $J_{R,\rm ABS}$ for 
dc biases with a superimposed microwave radiation described by 
$U_{L}(t)=U_{L}+U_{\rm r}\sin(\w_{\rm r}t)$, with $U_{\rm 
r}=0.05\w_{\rm r}$, $\w_{\rm r}=1.08$, and
$U_{L}=0.0,\;0.03,\;0.1,\;0.3$. The interplay between 
the ac Josephson effect and the resonant microwave driving leads to 
complicated TD patterns for small $U_{L}$. Increasing $U_{L}$ the 
life-time of the quasi ABS decreases resulting in 
a fast damping of the oscillations, see Fig. \ref{absfig} with 
$U_{L}=0.3$.

\subsection{Long atomic chains}

We consider a chain of $N+1=21$ atomic sites with onsite energy 
$\e_{C}=0$ and nearest neighbor hopping $t_{C}=1$, see Eq. 
(\ref{hccsp}), symmetrically coupled, $\G_{L}=\G_{R}=\G$, 
to superconducting electrodes with $|\D_{L}|=|\D_{R}|=\D$. 
In the limit of long chains one can prove that the current phase 
relation (at zero bias) is linear if $t_{C}=\G/2$.\cite{psc.2009,acz.2000}
This is the so called Ishii's sawtooth behavior\cite{i.1970} and is due to perfect AR.
To better visualize the MAR in the transient regime we therefore 
choose $t_{C}=\G/2$. In equilibrium there are $16$ ABS in 
the gap.
At time $t=0$ the system is driven out of 
equilibrium by a dc bias $U_{L}$ applied to lead $L$.

\begin{figure*}[t]
\centering
\includegraphics[width=1.\textwidth]{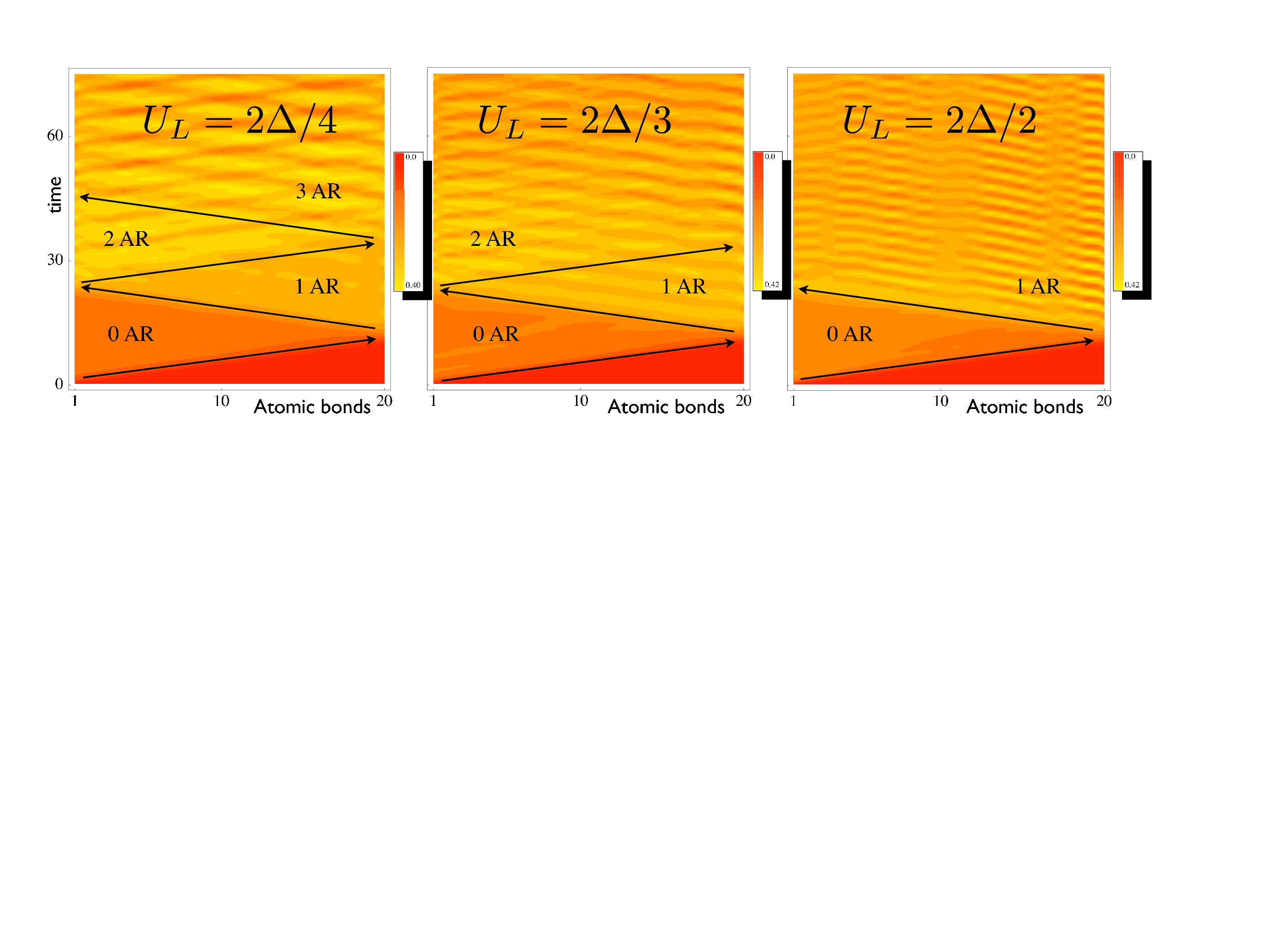}
\caption{TD picture of MAR. A chain of 21 atomic 
sites is symmetrically connected with $\G_{L}=\G_{R}=2t_{C}=2$ to two 
identical superconducting leads with $\D_{L}=\D_{R}=1$. A dc bias 
$U_{L}=2\D/n$, $n=4,\;3,\;2$, is applied to lead $L$ at time $t=0$.
The panels show the contour plots of the bond-current $J_{n,n+1}(t)$
across the atomic bonds of region $C$.
The results are obtained with a time-step $\d=0.05$, cut-off $\L=4$ 
and a number of scattering states $N_{p,L}=N_{p,R}=1232$.}
\label{perfect_AR}
\end{figure*}

In Fig. \ref{perfect_AR} we display the contour plot of the  
currents $J_{n,n+1}(t)$ along the bond $(n,n+1)$ of region $C$ as a 
function of time for different values of 
$U_{L}=2\D/4,\;2\D/3,\;2\D/2$. The MAR pattern is illustrated with 
black arrows. There is a clear-cut transient scenario during which 
electrons undergo $n$ AR before the ac Josephson regime sets in, with 
$n=U_{L}/2\D$. At every AR the current {\em increases} since the 
electrons are mainly reflected as holes and holes as electrons. The same numerical 
simulation in a normal system would have given a current in region 
1AR smaller than the current in region 0AR.

\begin{figure}[t]
\centering
\includegraphics[width=0.47\textwidth]{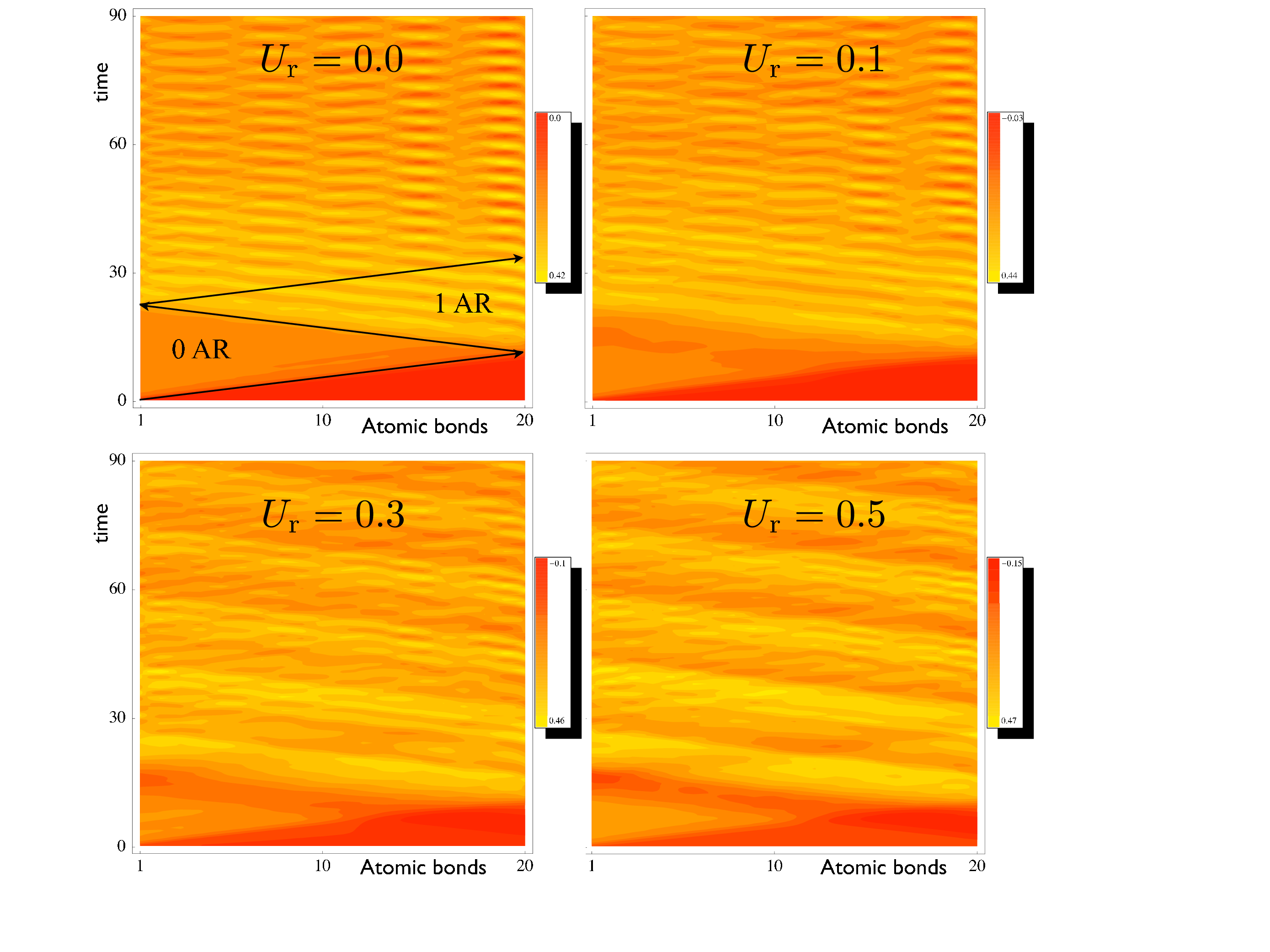}
\caption{Photon-assisted MAR in a chain of 21 atomic 
sites. The equilibrium parameters are the same as in Fig. \ref{perfect_AR}. 
An ac bias $U_{R}=U_{\rm r}\sin(\w_{\rm r}t)$ in lead $R$ is 
superimposed to a dc bias $U_{L}=0.8$ in lead $L$. 
The panels show the contour plots of the bond-current $J_{n,n+1}(t)$
across the atomic bonds of region $C$ for different values of $U_{\rm 
r}=0.0,\;0.1,\;0.3,\;0.5$ and $\w_{\rm r}=0.4$.
The results are obtained with a time-step $\d=0.05$, cut-off $\L=4$ 
and a number of scattering states $N_{p,L}=N_{p,R}=1232$.}
\label{photassistAR}
\end{figure}

For the same system parameters we also considered a dc bias 
$U_{L}=0.8$ for which the dominant scattering mechanism is the 3-rd 
order AR. The contour plot of the bond current is displayed in the 
top-left panel of Fig. \ref{photassistAR} and is similar to the case 
$U_{L}=2\D/3$ of Fig. \ref{perfect_AR}. A new scattering channel 
does, however, open if a microwave radiation of appropriate frequency 
is superimposed to $U_{L}$. We therefore applied an ac bias 
$U_{R}(t)=U_{\rm r}\sin(\w_{\rm r}t)$ to lead $R$ and choose 
$\w_{\rm r}$ to fulfill $2U_{L}+\w_{\rm r}=2\D$, i.e., 
$\w_{\rm r}=0.4$. In Fig. \ref{photassistAR} we report the contour plot of the 
bond-current for different values of $U_{\rm r}=0.0,\;0.1,\;0.3,\,0.5$.
At $U_{\rm r}\neq 0$ the right-going wave-front reduces its intensity 
just after crossing the bond 10 due to scattering against the 
left-going wave-front from lead $R$, see the characteristic 
$\l$-shape in the bottom-right panel.  When the right-going 
wave-front hits the right interface the bond current sharply 
increases. Furthermore, the larger is $U_{\rm r}$ the shorter is the 
transient regime. This can be explained as follows.
At large $U_{\rm r}$ the dominant 
scattering mechanism is the one in which an electron from lead $L$ 
and energy $U_{L}$ is reflected as a hole and at the same time 
absorbs a photon of energy $\w_{\rm r}$. The energy of the reflected
hole is $2U_{L}+\w_{\rm r}=2\D$, no extra AR are needed for charge 
transfer and the photon-assisted Josephson regime sets in.

\section{Conclusions and outlooks}
\label{conc}

In this paper we proposed a one-particle framework and a propagation 
scheme to study the TD response of UF-JNJ. 
By projecting the continuum Hamiltonian onto a suitable set of  
localized states we reduced the problem to the solution of a discrete 
system in which the electromagnetic field is described in terms of 
Peierls phases. The latter provide the basic quantities to construct 
a density functional theory of superconducting (and as a special 
case normal) systems. We proved that under reasonable conditions 
the TD bond current and pairing density of an interacting system 
driven out of equilibrium by Peierls phases $\g(t)$ can be reproduced 
in a system of noninteracting KS electrons under the influence of 
Peierls phases $\g'(t)$ {\em and} pairing field $\D'(t)$ and that 
$\g'(t)$ and $\D'(t)$ are unique.
We considered the KS system initially in equilibrium at given 
temperature and chemical potential when at time $t=0$ an external 
electromagnetic field is switched on. To calculate the response of 
the system at times $t>0$ we used a non-equilibrium formalism in which 
the normal and anomalous propagators are defined on an extended 
Keldysh contour that includes a purely imaginary (thermal) path going 
from 0 to $-i\b$.
We showed that the solution of the equations of motion for the 
NEGF are equivalent to first solve the 
static BdG equations and then the TD BdG equations. It is worth 
emphasizing that in TDSCDFT  the BdG equations do not follow from 
the BCS approximation and that their solution yields the exact 
bond-current and pairing density of an interacting system provided 
that the exact KS Peierls phases and pairing field are used.

For systems consisting of ${\cal N}$ superconducting leads in contact 
with a finite region $C$ and driven out of equilibrium by a 
longitudinal electric field a numerical algorithm is proposed. The 
initial eigenstates are obtained from a recent generalized wave-guide 
approach properly adapted to the superconducting case.\cite{spbc.2009}
The initial states are propagated in time using an embedded 
Crank-Nicholson algorithm which is norm-conserving, accurate up to 
second order in the time-step and that exactly incorporates 
transparent boundary conditions. The propagation scheme reduces to 
the one of Refs. \onlinecite{ksarg.2005,spc.2008} in the case of 
normal leads. 

The method described in this work allows for obtaining the TD current 
across an UF-JNJ and hence to follow the time evolution of several AR 
until the Josephson regime sets in. As a first calculation of these 
kind we explored in detail the popular single-level QD model in the 
weak and intermediate coupling regime. We demonstrated that the transient time 
increases with decreasing bias and provided a quantitative picture of 
the MAR. The rich structure of the transient regime is due to the 
ABS which play a crucial role in the relaxation process.
For dc pulses we showed that ABS can be exploited to generate {\em 
zero bias ac currents of tunable frequency}. Furthermore, irradiating 
the biased system with a microwave field of appropriate frequency the 
ABS give rise to a {\em long-living transient resonant effect}.
The transient regime increases also with the length of the junction. 
We considered one-dimensional atomic chains coupled to superconducting leads under dc 
and ac biases. 
Here we showed that in conditions of perfect AR there exists a 
clear-cut transient scenario for MAR. 
For biases $U_{L}=2\D/n$ the 
dominant scattering channel is the $n$-th order AR and 
the transient regime lasts for about $nN/v_{C}$ where $N$ is the 
length of the chain and $v_{C}$ the electron velocity at the Fermi 
level. Similar considerations apply to photon assisted MAR.
A more careful analysis of the transient regime is beyond the scope 
of the present paper. However such analysis is of utmost importance if 
the ultimate goal of superconducting nanoelectronics is to use 
these devices for ultrafast operations. 

The TD properties presented in this work have been obtained 
using rather simple, yet so far unexplored, models. 
A more sophisticated description of the Hamiltonian is, however, 
needed for a quantitative parameter-free comparison with experiments. 
Theoretical advances also involve the development of approximate 
functionals for the self-consistent calculation 
of the TD pairing potential and Peierls phases. 
Self-consistent calculations have so far been 
restricted to equilibrium S-D-S models  
with a point-like attractive interaction treated in the BCS 
approximation.\cite{mrgvly.1994,sps.2009,ma.1998,hma.1999}
For biased systems, however, the pairing potential and Peierls phases 
must be treated on equal footing and a first step in this direction 
would be the BCS 
approximation for the pairing field and the Hartree-Fock approximation for the 
Peierls phases. 
More difficult is the study of UF-JNJ in the Coulomb blockade regime 
for which electron correlations beyond Hartree-Fock  must be 
incorporated.

Finally, the approach presented in this work is not limited to two 
terminal systems. The coupling of the central region to a third 
normal lead, or gate, allows for controlling the Josephson current by varying 
the gate voltage.\cite{gpk.2008,wsz.1998,slsw.2000} 
These systems can be potentially used for fast switches and 
transistors,\cite{atn.1996,bmvwk.1999} and  a microscopic understanding of their 
ultrafast properties is therefore necessary to optimize their functionalities.

\appendix

\section{Calculation of the embedding matrices}
\label{embq}

Without loss of generality we include few layers of each lead in 
the explicitly propagated region $C$. Then, the embedding matrix 
$\ubQ^{(m)}_{\a}$ is zero everywhere except in the block of dimension 
$2N^{\a}_{\rm cell}\times 2N^{\a}_{\rm cell}$ which is connected to 
the $\a$ lead. Denoting with $\ubq^{(m)}_{\a}$ such 
non-vanishing block in $\ubQ^{(m)}_{\a}$ we have  
\begin{equation}
\ubq_{\a}^{(m)}=\ubt_{\a}\left[\frac{\left(\uno_{\a\a}-i\d\tilde{\ubH}_{\a\a}\right)^{m}}
{\left(\uno_{\a\a}+i\d\tilde{\ubH}_{\a\a}\right)^{m+1}}
\right]_{0.0}\ubt_{\a}^{\dag}\,,
\end{equation}
where the subscript $(0,0)$ denotes the first diagonal block (supercell with $j=0$) of the 
matrix in the square brackets. We notice that from Eq. 
(\ref{tildeaa}) the matrix $\tilde{\ubH}_{\a\a}$ is the same as the 
matrix $\ubH_{\a\a}(0)$ in Eq. (\ref{trdiag}) but with renormalized 
diagonal blocks $\tilde{\ubh}_{\a}=\ubh_{\a}-\mu\us_{\a}$.
In order to compute the $\ubq_{\a}^{(m)}$'s 
we introduce the generating matrix function
\begin{equation}
\ubq_{\a}(x,y)\equiv\ubt_{\a} \left[
\frac{1}{x\uno_{\a\a}+iy\d\tilde{\ubH}_{\a\a}}
\right]_{0,0}\ubt_{\a}^{\dag},
\label{q1def}
\end{equation}
which can also be expressed in terms of continued matrix fractions 
\begin{widetext}
\begin{eqnarray}
\ubq_{\a}(x,y)  &=& 
\ubt_{\a}
\mbox{$
        \frac{
	       \mbox{ {\normalsize $1$ } } 
	      }
	      {
	       \mbox{ {\normalsize $x\uno_{\a}+iy\d\tilde{\ubh}_{\a}+y^{2}\d^{2}\ubt_{\a}       
	                           \mbox{$
                                           \frac{
					         \mbox{ {\normalsize $1$ } } 
					        }
						{
						 \mbox{ {\normalsize 
						 $  x\uno_{\a} 
						 +iy\d\tilde{\ubh}_{\a}+y^{2}\d^{2}\ubt_{\a}
						                     \mbox{$
								             \frac{\mbox{ {\normalsize $1$ } } }{\ldots\ldots}
									   $}
						                     $ } 
						         {\normalsize 
							 $\ubt_{\a}^{\dag}$ }
				                       }
						}
			                 $}
	                           $ } 
		        {\normalsize $\ubt_{\a}^{\dag}$ }
	             }
	      }
      $}
\ubt_{\a}^{\dag}
\nonumber \\ 
&=&
\ubt_{\a}\frac{1}{x\uno_{\a}+iy\d\tilde{\ubh}_{\a}+y^{2}\d^{2}\ubq_{\a}(x,y)}\ubt_{\a}^{\dag}
\equiv \ubt_{\a}\ubp_{\a}(x,y)\ubt_{\a}^{\dag},
\label{qalpha}
\end{eqnarray}
\end{widetext}
where the last step is an implicit definition of $\ubp_{\a}(x,y)$.
The $\ubq_{\a}^{(m)}$'s are obtained from the generating matrix 
function as
\bea
\ubq_{\a}^{(m)}&=&\ubt_{\a}\frac{1}{m!}\left.\left[-\frac{\de}{\de x}+\frac{\de }{\de y}\right]^{m}
\ubp_{\a}(x,y)\right|_{x=y=1}\ubt_{\a}^{\dag}
\nonumber \\ 
&=&\ubt_{\a}\ubp_{\a}^{(m)}\ubt_{\a}^{\dag}.
\label{gen}
\eea
Using the identity $\frac{1}{m!}[-\frac{\de}{\de x}+\frac{\de}{\de y}]^{m}
\ubp_{\a}^{-1}(x,y)\ubp_{\a}(x,y)=0$, we derive the following recursive scheme
\bea
(\uno_{\a}\!+i\d\tilde{\ubh}_{\a})\ubp_{\a}^{(m)}
\!\!&=&\!\!(\uno_{\a}\!-i\d\tilde{\ubh}_{\a})\ubp_{\a}^{(m-1)} 
\nonumber \\ 
\!\!&-&\!\!\d^{2}\sum_{k=0}^{m}
(\ubq_{\a}^{(k)}\!+2\ubq_{\a}^{(k-1)}\!+\ubq_{\a}^{(k-2)})
\ubp_{\a}^{(m-k)}
\nonumber \\
\label{recur}
\eea
with $\ubp_{\a}^{(m)}=\ubq_{\a}^{(m)}=0$ for $m<0$. The above 
relation can be used to calculate $\ubq_{\a}^{(m)}$ provided that all 
$\ubp_{\a}^{(k)}$ are known for $k<m$. To obtain $\ubp_{\a}^{(0)}$ we 
can use Eq. (\ref{qalpha}) with $x=y=1$ in which the continued 
fraction is truncated after a number $N_{\rm level}$ 
of levels. Convergence can be easily checked by increasing $N_{\rm level}$.

\section{Calculation of the boundary term}
\label{bndtermcalc}

From Eq. (\ref{maintd}) we see that in order to propagate an 
eigenstate of $\ubH_{0}-\m\us$ we need to know the boundary term 
defined in Eq. (\ref{bndrterm}). The state $\F^{(0)}$ can be either a scattering 
state or an ABS. As shown in Section \ref{initeign} the 
projection onto lead $\a$ of a generic eigenstate with 
energy $E$ can be written as a linear combination of states of the 
form
\be
\F^{\a}_{k}(m=s,j,\a)=Z_{k}^{\a}(s)e^{ikj},
\label{genproj}
\ee
where the amplitudes $Z_{k}^{\a}$ satisfies the eigenvalue equation
\be
\left(\ubh_{\a}+\ubt_{\a}e^{ik}+\ubt_{\a}^{\dag}e^{-ik}-\m\us_{\a}
\right)Z_{k}^{\a}=EZ_{k}^{\a}.
\label{eigeq}
\ee
In the following we show how to compute the action of the operator
$\tilde{\ubH}_{C\a}(0)
\ubg_{\a\a}^{m}\left(\uno_{\a\a}+\ubg_{\a\a}\right)$ on 
$\F^{\a}_{k}$. 
We define the Nambu vector in region $C$
\bea
\F_{C,k}^{\a(m)}&\equiv& \tilde{\ubH}_{C\a}(0)
\ubg_{\a\a}^{m}\left(\uno_{\a\a}+\ubg_{\a\a}\right)
\F^{\a}_{k}
\nonumber \\
&=&2\tilde{\ubH}_{C\a}(0)
\frac{\left(\uno_{\a\a}-i\d\tilde{\ubH}_{\a\a}\right)^{m}}
{\left(\uno_{\a\a}+i\d\tilde{\ubH}_{\a\a}\right)^{m+1}}
\F^{\a}_{k},
\label{bdrtm}
\eea
from which the boundary term can easily be extracted by taking the 
appropriate linear combination of the $\F_{C,k}^{\a(m)}$ and then
multiplying by $-i\d\ubcalz_{\a}^{(m)}$, see Eq.  (\ref{bndrterm}). 
Since region $C$ includes few layers of the leads
the vector $\F_{C,k}^{\a(m)}$ is zero 
everywhere except for the components corresponding to orbitals 
in contact with lead $\a$. If we call $\f_{C,k}^{\a(m)}$ the 
vector with such components from Eq. (\ref{bdrtm}) we can write
\be
\f_{C,k}^{\a(m)}
=2\ubt_{\a}\left[
\frac{\left(\uno_{\a\a}-i\d\tilde{\ubH}_{\a\a}\right)^{m}}
{\left(\uno_{\a\a}+i\d\tilde{\ubH}_{\a\a}\right)^{m+1}}
\F^{\a}_{k}\right]_{j=0}\equiv 2\ubt_{\a}V_{k}^{\a(m)},
\label{bdrtm2}
\ee
where the subscript $j=0$ in the square brackets denotes the vector 
of dimension $2N^{\a}_{\rm cell}$ with 
components given by the projection of 
the full vector onto the first ($j=0$) supercell. As for the embedding 
matrices we introduce the generating function 
\be
V_{k}^{\a}(x,y)=
\left[
\frac{1}
{x\uno_{\a\a}+iy\d\tilde{\ubH}_{\a\a}}
\F^{\a}_{k}\right]_{j=0}
\ee
from which the $V_{k}^{\a(m)}$ are obtained via multiple 
derivatives 
\be
V^{\a(m)}_{k}=\frac{1}{m!}\left.\left[-\frac{\de}{\de x}+\frac{\de}{\de y}\right]^{m}
V_{k}^{\a}(x,y)\right|_{x=y=1}.
\label{vgen}
\ee
The generating function can be obtained as follows. 
Taking $\F^{\a}_{k}$ as in Eq. (\ref{genproj}) and exploiting the 
property in Eq. (\ref{eigeq}) it is easy to realize that
\begin{equation}
\left[\tilde{\ubH}_{\a\a}\F^{\a}_{k}\right]_{j}=
(E-\d_{j,0}e^{-ik}\ubt_{\a}^{\dag})\left[\F^{\a}_{k}\right]_{j},
\label{siee}
\end{equation}
where the subscript $j$ denotes the vector 
of dimension $2N^{\a}_{\rm cell}$ with 
components given by the projection of 
the full vector onto the $j$-th supercell.
Then, multiplying the Dyson identity
\begin{equation}
\frac{1}
{x\uno_{\a\a}+i\d y\tilde{\ubH}_{\a\a}}=\frac{1}{x}
-\frac{iy\d}{x}\frac{1}
{x\uno_{\a\a}+iy\d\tilde{\ubH}_{\a\a}}\tilde{\ubH}_{\a\a}
\label{dysv}
\end{equation}
on the right by $\F^{\a}_{k}$, using Eq. (\ref{siee}) 
and solving for $V^{\a}_{k}(x,y)$ we obtain the following result
\begin{equation}
V^{\a}_{k}(x,y)=\frac{1+iy\d e^{-ik}\ubp_{\a}(x,y)\ubt_{\a}^{\dag}}{x+iy\d E}
Z^{\a}_{k},
\label{s1st}
\end{equation}
where $\ubp_{\a}(x,y)$ is the generating function defined in Eq. 
(\ref{qalpha}). The quantity $V^{\a(m)}_{k}$ can now be obtained 
from Eq. (\ref{vgen}) and reads
\bea
V^{\a(m)}_{k}&=&
\frac{(1-i\d E)^{m}}{(1+i\d E)^{m+1}}Z^{\a}_{k}+i\d e^{-ik}
\sum_{n=0}^{m}
\frac{(1-i\d E)^{m-n}}{(1+i\d E)^{m-n+1}}
\nonumber \\
&\times&\left(
\ubp_{\a}^{(n)}+\ubp_{\a}^{(n-1)}
\right)\ubt_{\a}^{\dag}Z^{\a}_{k}.
\eea
This conclude the calculation of the boundary term.

\bibliographystyle{apsrev}

\end{document}